\documentclass[aps,prd,twocolumn,superscriptaddress,nofootinbib,10pt,preprintnumbers]{revtex4-2}

\pdfoutput=1
\usepackage[utf8]{inputenc}
\usepackage{graphicx,amsmath,amsfonts,amssymb,aas_macros,slashed}
\usepackage{bbold, bm}
\usepackage{datetime}
\usepackage{numprint}
\usepackage{enumitem}
\usepackage{graphicx,color,amsmath}
\usepackage{hyperref}
\usepackage{mathtools}
\usepackage{booktabs}
\usepackage{pifont}%
\newcommand{\cmark}{\ding{51}}%
\newcommand{\xmark}{\ding{55}}%

\usepackage{rotating}
\allowdisplaybreaks

\newcommand{\cm}{\ensuremath{\mathrm{cm}}}

\newcommand{\MeV}{\ensuremath{\mathrm{MeV}}}

\newcommand{\tmu}{{\tilde \mu}}

\newcommand{\hS}{{\hat S}} 
\newcommand{\hs}{{\hat s}}

\usepackage{pgfplots}
\usepackage{bm}

\makeatletter
\newcommand{\thickhline}{%
    \noalign {\ifnum 0=`}\fi \hrule height 1pt
    \futurelet \reserved@a \@xhline
}
\newcolumntype{"}{@{\hskip\tabcolsep\vrule width 1pt\hskip\tabcolsep}}
\makeatother

\begin{document}
\preprint{UWThPh 2023-24}

\title{On the minimal mass of thermal dark matter\\ and  the viability of millicharged particles affecting 21cm cosmology}

\author{Xiaoyong Chu}
\affiliation{Institute of High Energy Physics, Austrian Academy of Sciences, Nikolsdorfergasse 18, 1050 Vienna, Austria}

\author{Josef Pradler}
\affiliation{Institute of High Energy Physics, Austrian Academy of Sciences, Nikolsdorfergasse 18, 1050 Vienna, Austria}
\affiliation{University of Vienna, Faculty of Physics, Boltzmanngasse 5, A-1090 Vienna, Austria}

\begin{abstract}
Thermal freeze-out offers an attractive explanation of the dark matter density free from fine-tuning of initial conditions. For dark matter with a mass below tens of MeV, photons, electrons, and neutrinos are the only available direct Standard Model annihilation products. Using a full three-sector abundance calculation, we determine the minimal mass of dark matter, allowing for an arbitrary branching into electrons/photons and neutrinos that is compatible with current cosmological observations. The analysis takes into account the heat transfer between the various sectors from annihilation {\it and} elastic scattering, representing the first fully self-consistent analysis that tracks the respective sectors' temperatures. We thereby provide accurate thermal annihilation cross sections, particularly for velocity-dependent cases, and deduce the sensitivity of current and  upcoming CMB experiments to MeV thermal dark matter. In the latter context, we also establish the fine-tuned parameter region where a tiny admixture of neutrinos in the final states rules in MeV-scale $p$-wave annihilating DM into electrons. 
Finally, we show that a sub-\% millicharged dark matter with an interaction strength that interferes with 21~cm cosmology is still allowed when freeze-out is supplemented with annihilation into neutrinos. For all cases considered, we provide concrete particle physics models and supplement our findings with a discussion of other relevant experimental results.
\end{abstract}

\maketitle

\section{Introduction}

Weakly interacting massive particles (WIMPs) make for attractive dark matter (DM) candidates: the combination of electroweak-scale mass and interactions---in strength reminiscent of the weak interactions in the Standard Model (SM)---allow for a broad experimental and observational program in their search. In this quest, the parameter space that predicts the correct relic abundance provides an important experimental target. In the early universe, WIMPs come into thermal equilibrium with the SM, and their non-relativistic chemical decoupling allow for an understanding of their density that is free from initial conditions.

The to-date absence of new physics at the electroweak scale, however,  has motivated efforts to experimentally probe an increased range in DM mass. 
Particularly significant advances, both theoretically and experimentally, have allowed us to push the sensitivity of directly detecting dark matter in the laboratory below the 100~MeV mass scale. 
This recently gained sensitivity is, on the other hand, not easily matched with cosmologically compatible models of thermal DM relics.
The number of available annihilation channels for thermally regulating the DM abundance reduces drastically, while the demand on the size of the cross section increases. At the same time, freeze out happens close to the  highly non-trivial epochs of neutrino decoupling and electron-positron annihilation. 
A thorough calculation of the relic density must hence relate to a three sector system, the electromagnetic sector (photons, electrons), neutrinos and DM. When this system is appropriately solved, it not only predicts the relic density but also the temperature ratio of neutrinos to photons, which, by itself is an important and sensitive cosmological observable and often in tension when considering a MeV-mass~DM.

Previous treatments of thermal MeV-scale DM have mostly assumed instantaneous
neutrino decoupling~\cite{Ho:2012ug,Boehm:2012gr,Steigman:2013yua,Boehm:2013jpa,Green:2017ybv}. More recently, 
systematic efforts toward a full treatment of MeV-scale thermal DM decoupling and a study of cosmological observables were made in~\cite{Escudero:2018mvt} and in~\cite{Depta:2019lbe, Sabti:2019mhn,An:2022sva}. These works account for the energy transfer from the dark to the SM sector from annihilation with the aim to precisely predict~$N_{\rm eff}$ and/or light element abundances. 
The main caveat in the above-mentioned works is that it had to be assumed that DM stays in thermal equilibrium with either photons or neutrinos, while classical Maxwell-Boltzmann statistics and an annihilation cross section independent of DM mass had to be adopted.  

An important step towards a fully self-consistent treatment of the problem that allows for arbitrary branching ratios into neutrinos and photons/electrons was provided by the authors of Ref.~\cite{Chu:2022xuh}. There, the three-sector problem  is formulated in such a way that it becomes computationally feasible to solve the coupled set of Boltzmann equations over a great numerical range of reaction rates while ensuring fulfillment of the detailed balancing conditions. Moreover, for the first time,  it became possible to include the energy transfer between the sectors originating from the number-conserving {\it elastic} scattering processes. 

The purpose of this paper is to follow up on the introduced methodology in~\cite{Chu:2022xuh} and provide concrete examples of MeV-scale DM decoupling. 
For generic WIMP DM, the canonical thermal cross section is $\langle \sigma v \rangle \approx 3\times 10^{-26}\ \cm^3/{\rm s}$, independent of the WIMP mass~\cite{Kolb:1988aj,Gondolo:1990dk}. However, as is now well known, this value is subject to changes, particularly for light DM
\cite{Steigman:2012nb, Bringmann:2020mgx}.
In this work, we compute the exact value of the required thermally averaged cross section that provides the correct relic density for a set of relative branching ratios of the annihilation channels into neutrinos vs.~electrons, for $s$- and $p$-wave annihilation cross sections in a DM mass regime where freeze-out overlaps with neutrino decoupling. By dialing through the branching ratios  we further explore the minimal DM mass that is compatible with the cosmic microwave background (CMB) measurements and if a careful partition of branching ratios allows for an avoidance of this constraint while simultaneously maintaining the successful big bang nucleosynthesis (BBN) predictions.

In a second part, we apply our methods to millicharged dark states.
The existence of such particles can have far-reaching consequences for phenomenology, astrophysics, and cosmology.  
Because the non-relativistic elastic scattering on baryons is enhanced with relative velocity $v$ as $v^{-4}$, such relic can also induce a cooling of the baryonic gas in the post-recombination Universe when DM is at its coldest temperature. This has been shown to affect the expected cosmological neutral hydrogen 21\,cm absorption signal at the epoch of the cosmic dawn~\cite{Tashiro:2014tsa,Munoz:2018pzp,Barkana:2018qrx}{and may, additionally, impact structure formation~\cite{Driskell:2022pax}.}  Although the observational status {of the 21~cm signal} is unclear, with a putative detection of a global absorption feature by  EDGES~\cite{Bowman:2018yin} but not confirmed by SARAS3~\cite{Singh:2021mxo}, the prospect of probing dark sector properties through 21\,cm cosmology is exciting. 

The proposal faces very stringent limits from direct detection, fixed target experiments, {high-redshift observables}\,\cite{Barkana:2018qrx,Fraser:2018acy,Berlin:2018sjs,Kovetz:2018zan,Slatyer:2018aqg,Wadekar:2019mpc,Creque-Sarbinowski:2019mcm,Liu:2018uzy,Liu:2019knx,Emken:2019tni,Buen-Abad:2021mvc,SENSEI:2023gie}, among other limits, and, jointly, they require millicharged DM to constitute only a sub-percent fraction of the total DM abundance and be situated in a particular corner of parameter space: MeV-scale mass and a millicharge $Q\sim 10^{-5}-10^{-4}$.%
\footnote{Additional constraints, in particular cosmological ones~\cite{Vogel:2013raa}, arise when the millicharged DM is realized through an ultralight kinetically mixed dark photon. In this work, we do not consider this further possibility.}
Even satisfying all those constraints, the cosmic viability remains questionable because the thermalization and guaranteed annihilation of these particles into electrons lowers the $N_{\rm eff}$ value to unacceptable values~\cite{Creque-Sarbinowski:2019mcm}.
In this work, we revisit the possibility of millicharged DM by supplying its interactions with an additional annihilation channel into neutrinos. The joint annihilation into electrons and neutrinos harbors the potential of alleviating the $N_{\rm eff}$ constraint, hence widening or ruling in this possibility. This extension requires a three-sector treatment and  is therefore a perfect application of our here-developed methodology.

The paper is organized as follows. 
In Sec.~\ref{sec:sandpwave} we produce the values of the thermal DM cross sections for DM masses below 20~MeV and find the minimum cosmologically compatible mass. 
In Sec.~\ref{sec:millicharged} we compute the $N_{\rm eff}$ and nucleosynthesis predictions for millicharged DM when it is supplied with neutrino interactions in the coupling range where DM-baryon interactions affect the global 21~cm signal. We conclude in Sec.~\ref{sec:conclusions}. Several appendices provide some semi-analytical solutions to the thermal DM cross sections, as well as the calculational results that go into the solution of the Boltzmann equations.

 \begin{figure}[t]
\begin{center}
\includegraphics[width=\textwidth,viewport=0 710 1300 1100,clip=true]{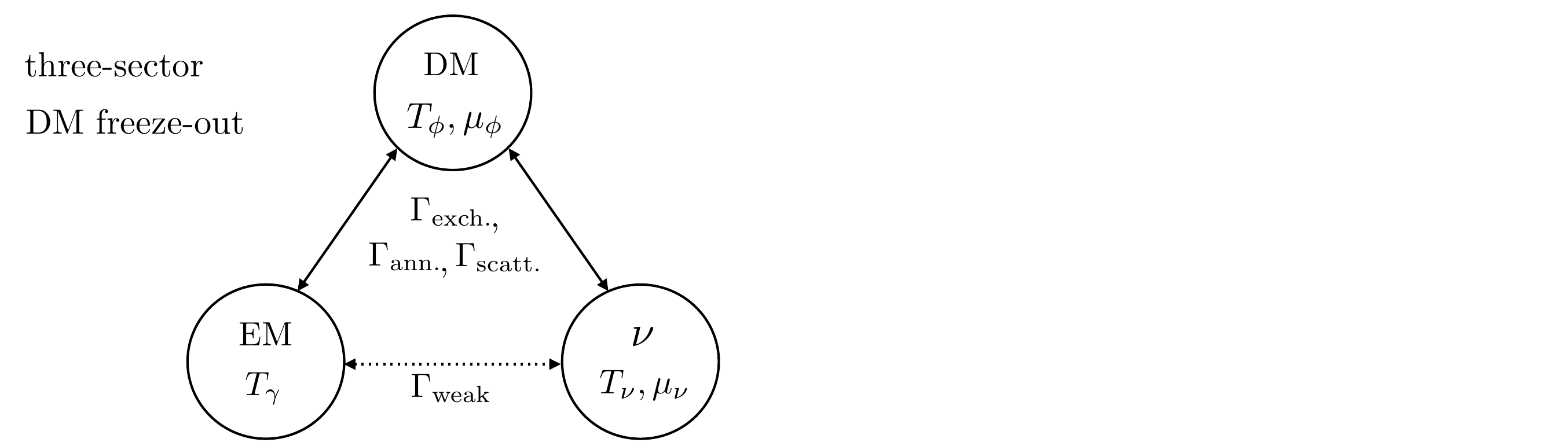}
\caption{Schematic depiction of the three coupled sectors (EM, $\nu$, DM ``$\phi$'' or ``$\chi$'') with the respective variables we solve for: the temperatures $T_\gamma$, $T_\nu$, $T_\phi$ and chemical potentials $\mu_\nu$ and $\mu_\phi$ in the neutrino and DM sector. The sectors are kept in contact by various rates: SM weak interactions  $\Gamma_{\rm weak}$, DM annihilation into SM states $\Gamma_{\rm ann.}$, energy exchange from annihilation $\Gamma_{\rm exch.}$,  and elastic scattering $\Gamma_{{\rm scatt.}}$.
}
\label{fig:scheme}
\end{center}
\end{figure}

\section{Thermal cross sections and \boldmath$N_{\rm eff}$}
\label{sec:sandpwave}

The objective of this section is to provide a precise value of the thermally averaged cross section in the situation that annihilation  occurs during or in the vicinity of the epoch of neutrino decoupling. This necessitates a simultaneous solution for three-sectors: the electromagnetic (``EM'') one comprised of electrons, positrons and photons,  the SM neutrino  (``$\nu$'') one, and the dark matter sector (``$\phi$'' or ``$\chi$'').%
\footnote{If right-handed neutrinos are also light, and DM annihilates into both left- and right-handed neutrinos, our method  can be applied to the evolution of a different three-sector scenario: ``DM'', ``SM'', and ``right-handed $\nu$;'' see \cite{Abazajian:2019oqj} for a two-sector case. }

In a previous work we have developed the methodology~\cite{Chu:2022xuh} for such treatment. It is based on a reformulation that makes detailed balancing numerically manifest for quantum statistics, together with a factorization of neutrino and DM chemical potentials in the respective collision terms. 
A pictorial representation of the three-sector problem is given in Fig.~\ref{fig:scheme}.
The dynamics is governed by a number of rates, where the most familiar ones are the rate of weak interactions, 
$\Gamma_{\rm weak}  \equiv  n_e G_F^2 T_\gamma^2 $, determining neutrino decoupling and the total DM annihilation rate
$\Gamma_{\rm ann.}  \equiv n_\phi \langle \sigma_{\rm ann.} v \rangle $ controlling the DM number density and determining the point of chemical decoupling from the SM bath; $G_F$ is the Fermi constant, $T_\gamma$ the photon temperature and  $n_{e/\phi}$
is the electron/DM number density.
A related important rate controlling the temperature evolution of the various sectors is the energy exchange rate $\Gamma_{{\rm exch.},i}  \equiv n_\phi^2  \langle \sigma_{{\rm ann.}, i} v \delta E\rangle /\rho_{i} $ between DM and sector $i\in\{\rm EM,\ \nu\}$. Here, the thermal average is weighted by the energy $\delta E$ that is transfer between the sectors.
For example, when $\Gamma_{\rm weak}<H$ but  $\Gamma_{{\rm exch.},i} >H$, $T_\nu =T_\gamma$ remains possible even after neutrino decoupling in a standard cosmology. Tracking the energy exchange is hence of crucial importance in the determination of  $N_{\rm eff}$. Finally, energy may also be transferred by number conserving {\it elastic} scatterings with a rate given by
$\Gamma_{{\rm scatt.}i} \equiv n_\phi n_i \langle \sigma_{\rm scatt.}^{\phi, i}  v \,\delta E\rangle / \rho_i$ and where the typical energy transfer is of the order of the temperature difference between the sectors.
 We are able to include this channel across the enormous dynamical range in particle densities and rate efficiencies. We refer the reader to~\cite{Chu:2022xuh} for a detailed discussion of the rates and the ensuing sequence of DM decoupling. 

\subsection{{Parametrization and existing constraints}}

{In this work we consider the thermal histories of complex scalar DM~$\phi$ and Dirac fermion DM~$\chi$ at the MeV mass-scale, annihilating into electromagnetically interacting particles and neutrinos with respective branching ratios ${\rm B}_{\rm EM}$ and ${\rm Br}_\nu$, with ${\rm B}_{\rm EM} + {\rm Br}_\nu$ =1.}
The total annihilation cross section times the M{\o}ller velocity $v_M$ is parametrized in the usual non-relativistic expansion of relative velocity $v_{\rm rel}$,%
\footnote{For velocity-dependent  annihilation it can make a difference at order $v_{\rm rel}^2$ if a velocity expansion of the Lorentz-invariant product $\sigma v_{M}$ or of $\sigma v_{\rm rel}$ is considered; see~\cite{Boehm:2020wbt} for a concrete $p$-wave example.}
\begin{align}
    \sigma_{\rm ann} v_{M} = a + bv_{\rm rel}^2 + \mathcal {O}(v_{\rm rel}^4)
\end{align}
A non-relativistic thermal average then yields $\langle \sigma v_M \rangle = a + 6 b/x + \dots$\, where we have used that $\langle v_{\rm rel}^2 \rangle  \simeq  6T_\phi /m_\phi =  6/x$. 
For orientation, the canonical values for a Majorana fermion with mass above 10~GeV
are $a \approx 2\times 10^{-26}\,\cm^3/{\rm s} = 1.7 \times 10^{-15}\ \MeV^{-2}$ for pure $s$-wave  annihilation ($b=0$) and $b \approx 1.5\times 10^{-25}\,\cm^3/{\rm s}   =1.3 \times 10^{-14}\ \MeV^{-2} $ for pure $p$-wave annihilation $(a=0)$. 

%
%
%
%
%

\begin{figure*}[t]
\begin{center}
\includegraphics[width=0.50\textwidth]{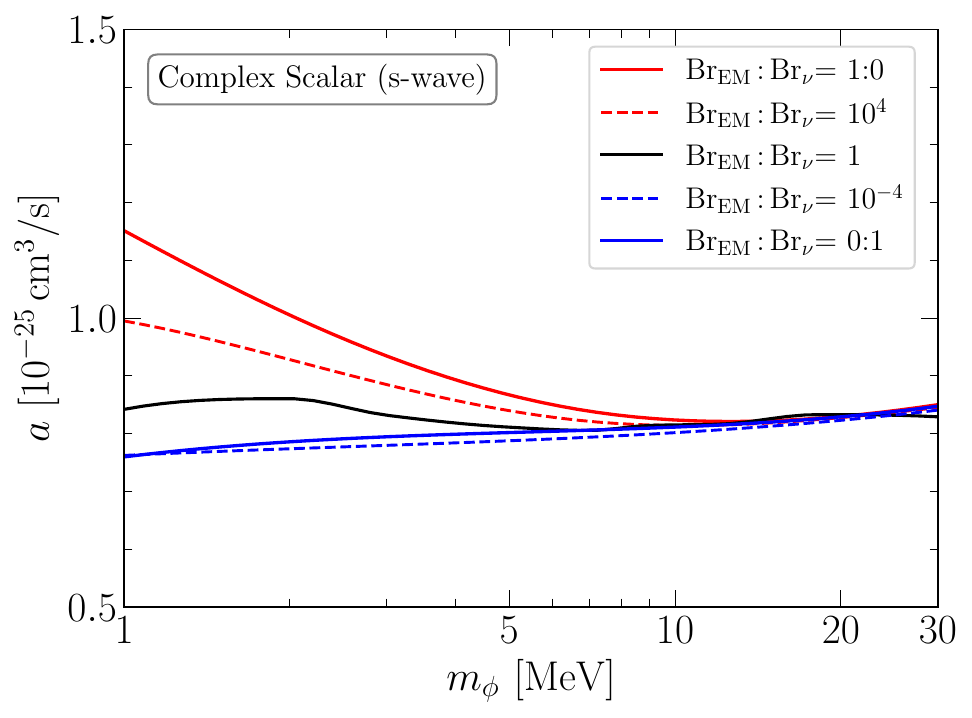}
\includegraphics[width=0.485\textwidth]{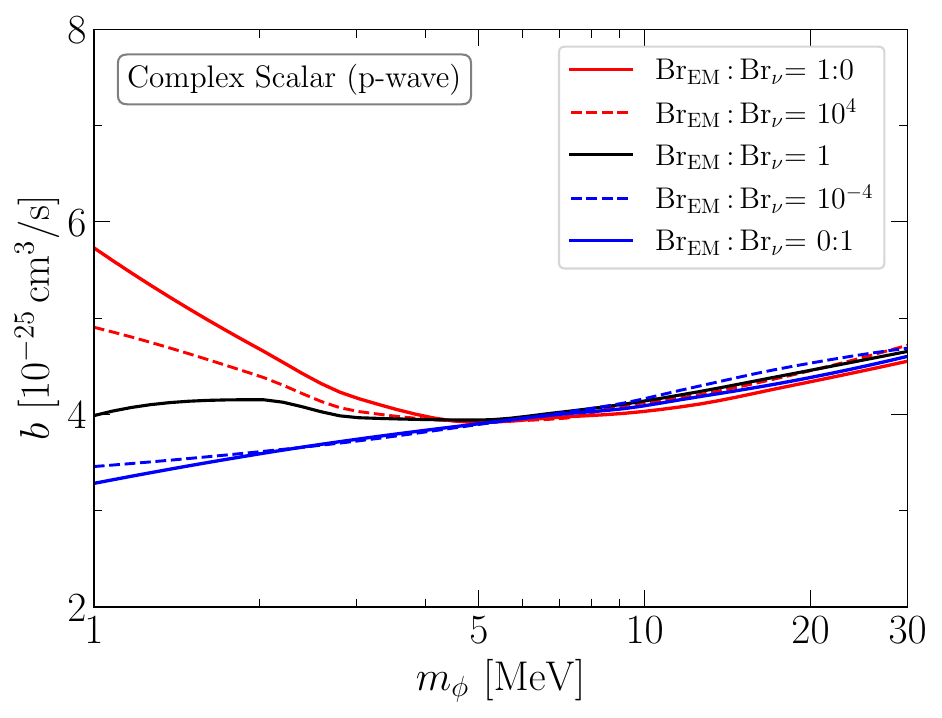}
\end{center}
\caption{Parameters of annihilation cross sections for thermal freeze-out of a complex scalar $\phi$ as a function of DM mass $m_\phi$ that yields the correct relic DM density from numerically solving the coupled three-sector system. The left (right) panel is for $s$-wave ($p$-wave) annihilation. Curves for different branching ratios as labeled are shown with different colors and dashing.   
}
\label{fig:thermal_xsec}
\end{figure*}

{
The annihilation cross section into the EM sector faces stringent limits from cosmology.}
The strongest constraints on $s$-wave MeV-mass DM come from Planck observations of the CMB, suggesting $a \times {\rm BR}_{\rm EM}   \lesssim (3\text{--}4)\times 10^{-30}\,{\rm cm}^3/{\rm s}$~\cite{Slatyer:2015jla}, where ${\rm BR}_{\rm EM}$ is the annihilation branching ratio into the EM sector; see also~\cite{Bartels:2017dpb, Liu:2018uzy,Wadekar:2021qae} for other indirect detection limits and prospects thereof. In practice, the CMB constraint demands that a thermal $s$-wave freeze-out in our considered mass range must obey ${\rm BR}_{\rm EM} \le  10^{-4}$, although we will consider arbitrary values of ${\rm BR}_{\rm EM}$ for completeness below. In contrast, due to the velocity suppression factor, $p$-wave freeze-out with $b \times {\rm BR}_{\rm EM}  \sim  10^{-25}\,{\rm cm}^3/{\rm s}$ is currently not constrained by CMB and low-redshift indirect searches; see e.g. \cite{Boudaud:2016mos, Boudaud:2018oya}.  
Constraints on DM annihilation into neutrinos are significantly weaker. For the considered DM mass range below $20~\MeV$, values of $a \times {\rm BR}_{\nu} \lesssim (1\text{--}100)\times 10^{-24}\,{\rm cm}^3/{\rm s}$, are allowed by the combination of Borexino, KamLAND and Super-Kamiokande~\cite{Arguelles:2019ouk}. Thus it leaves enough parameter space for a $s$-wave thermal freeze-out, let alone a $p$-wave one. 

{Before proceeding to present the results, we point out that taking into account the elastic scattering between DM and electrons/neutrinos necessitates the specification of a model, as there is no universal correspondence between annihilation and elastic scattering. The concrete models that we use are specified below in Sec.~\ref{sec:models} and in App.~\ref{app:DiracF} as well as in the companion paper~\cite{Chu:2022xuh}. 
The induced model-dependence is nevertheless relatively mild. For $s$-wave annihilation, the elastic scattering is irrelevant and the model dependence in fact drops out. For $p$-wave annihilation, the models are chosen to be as minimal as possible: they assume that the same heavy mediator that enables the annihilation is also responsible for the elastic scattering. In this sense, also the $p$-wave results are generic.%
\footnote{{One may of course construct rather special cases, by tuning parameters or introducing multiple mediators, e.g.~DM dominantly annihilating into electrons while being kinetically coupled to neutrinos. This is beyond the scope of this work.}}
}

\subsection{Thermal cross section values}

After solving the whole set of Boltzmann equations, we show in Fig.~\ref{fig:thermal_xsec}  the required thermal annihilation cross sections for $\phi$ from the joint solution for the coupled sectors. The left panel is for $s$-wave annihilation. For $m_{\phi} \ge 25\,$MeV, DM freeze-out happens well before  neutrino-electron decoupling, so varying the branching ratios has little impact. In this region, reducing the DM mass requires a ``closer-to-relativistic'' freeze-out, or, equivalently, a smaller value of $x_{\rm f.o.} \equiv m_\phi/T_{\rm f.o.}$ where $T_{\rm f.o.}$ is the chemical decoupling temperature.
Therefore, the correct relic abundance requires the canonical annihilation cross section to become smaller with decreasing $\phi$-mass, in accordance with the well-known relation $Y_{\phi} \propto x_\text{f.o.}/\langle \sigma_{\rm ann} v_M \rangle$~\cite{Kolb:1990vq}.

For  $m_{\phi} \lesssim 10\,$MeV, the $s$-wave  freeze-out goes through a period where the EM sector is being reheated by electron-positron annihilation. Because of the elevated photon temperature relative to the neutrino temperature, DM that is dominantly coupled to the EM sector enjoys a higher abundance relative to DM that is dominantly coupled to neutrinos. 
Consequently, in order to yield the correct relic abundance for DM coupled to electrons, annihilation should last longer than compared to without reheating,  requiring a larger annihilation cross section.
In the opposite case that DM dominantly annihilates into neutrinos, the EM sector affects freeze-out only indirectly through the evolution of the effective degrees of freedom. The latter is, however, of little importance because  electron-positron annihilation happens in an entropy-conserving manner ($S=s a^3=const$) so that the final DM yield $Y_\phi = n_\phi a^3/S$ is affected only mildly through the term $\frac{1}{3}\frac{d\ln g_s}{d\ln x}$ in the Boltzmann equation; see Eq.~\eqref{eq:Boltzmann} in the Appendix.%
\footnote{
A similar effect also arises from DM annihilation, 
which may preferably reheat the EM or the neutrino sector thereby changing the entropy degrees of freedom  with respect to the temperature of photons or neutrinos, respectively. In our code, this is taken into account.
} 
This explains Fig.~\ref{fig:thermal_xsec} where the annihilation cross section into electrons has a stronger trend than the annihilation cross section into neutrinos. 

The middle ground between the two extreme branching ratios is more involved, and is sensitive to the energy transfer among the three sectors. Concretely, the presence of a dark sector now causes two effects. One is it may maintain the kinetic equilibrium of neutrino and EM sectors, \textit{i.e.}, $T_\nu = T_\gamma$, via the  energy exchange between EM and neutrino sectors  mediated by DM interactions even after neutrinos decouple from electrons.  This effect thus tends to increase $T_\nu/T_\gamma$  in comparison to a standard cosmology after electron decoupling. The other effect is DM annihilation after EM-neutrino kinetic decoupling, which may increase or decrease $T_\nu/T_\gamma$, depending on the annihilation branching ratio ${\text{Br}_{\rm EM}:\text{Br}_\nu}$. For most of the parameter space, the first effect dominates over the second one. More interestingly, we illustrate in the next subsection that one may tune the value of ${\text{Br}_{\rm EM}:\text{Br}_\nu}$ to make the two effects cancel with each other, bringing the final $T_\nu/T_\gamma$ ratio close to its standard cosmology value~$0.7164$.  Note that the DM-induced energy transfer is dominated by DM pair creation/annihilation  in the $s$-wave case, and the kinetic decoupling of neutrinos from the EM sector is mainly sensitive to the product ${\text{Br}_\nu\,\text{Br}_{\rm EM}}$.

\begin{figure*}[t]
\begin{center}
\includegraphics[width=0.49\textwidth]{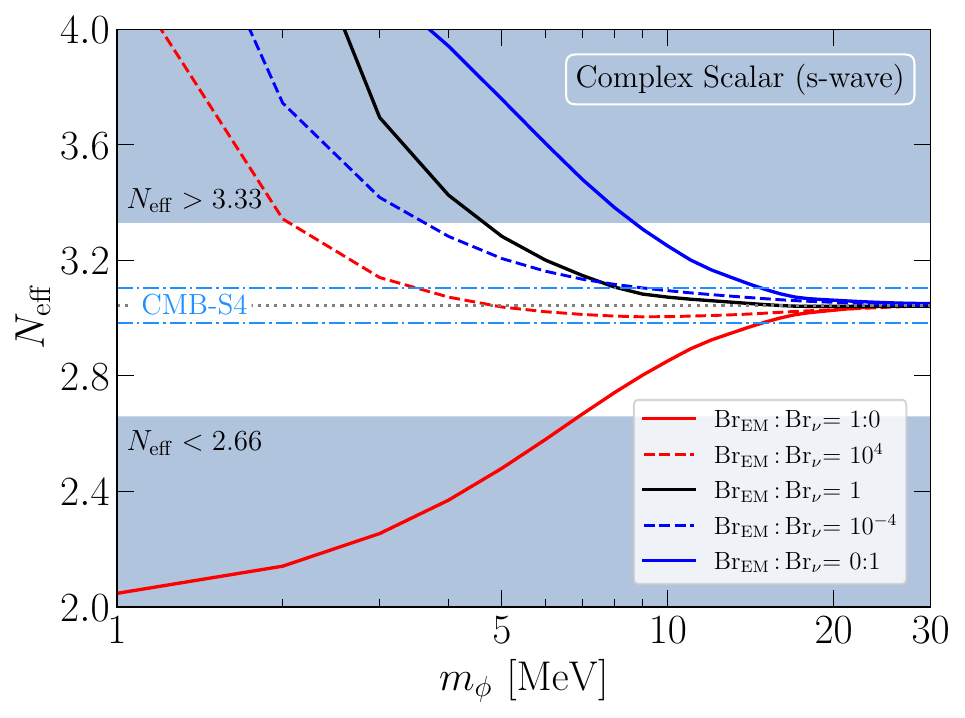}
\includegraphics[width=0.49\textwidth]{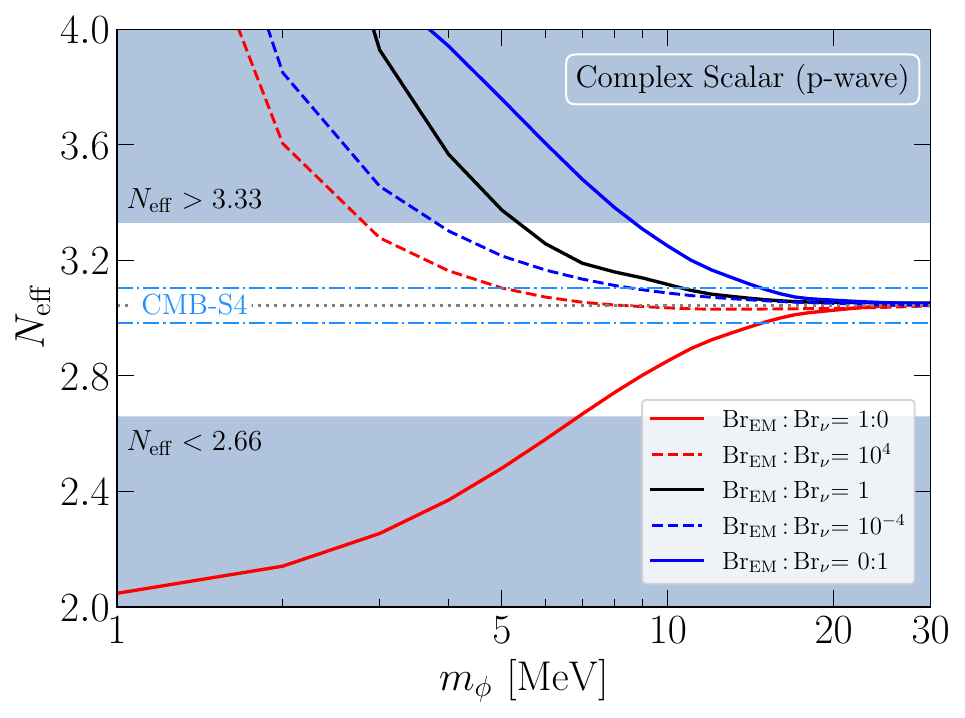}
\end{center}
\caption{\textit{Left panel}: 
$N_{\rm eff}$ values obtained for complex scalar DM
with the thermal annihilation cross-section that matches the DM abundance. The left (right) panel is for $s(p)$-wave annihilation. The calculation accounts for the energy transfer from annihilation {\it and} elastic scattering among the  various sectors. Shaded regions are excluded from combining Planck\,$+$\,BAO data, while the CMB-S4  sensitivity (standard cosmology value $N_{\rm eff}= 3.044$)  illustrated by the dot-dashed (dotted) horizontal lines. 
}
\label{fig:thermal_Neff}
\end{figure*}

The right panel of Fig.~\ref{fig:thermal_xsec} shows the result for $p$-wave annihilation. It generalizes our previous work~\cite{Chu:2022xuh} where we only entertained a fixed value ${\text{Br}_{\rm EM}:\text{Br}_\nu} = 2:3$. Here, we consider a broader variety in analogy to the $s$-wave case, with  similar features observed.  As established in \cite{Chu:2022xuh}, taking into account elastic scattering is crucial in the $p$-wave case. The reason is that  elastic scattering is able to maintain the kinetic coupling between DM and neutrinos (or electrons, depending on the branching ratios) for a longer period after DM freeze-out.
This leads to a mild heating of DM particles. In other words, elastic scattering affects the average DM velocity, and hence feeds into the calculation of the velocity-dependent cross section from the freeze-out point onward.

Moreover, in the $p$-wave case the effective DM annihilation cross section quickly decreases with time, so that the final DM abundance only marginally depends on the period $x \gg x_{\rm f.o.}$ (little residual annihilation). In contrast, for the $s$-wave case, residual DM annihilation decreases the abundance by more than 10\% from $x= 100$ until the CMB epoch. As a result, in the $p$-wave case, the EM sector reheating by electron annihilation only starts to play a role for somewhat lighter DM compared to the $s$-wave case (left panel of Fig.~\ref{fig:thermal_xsec}).  Meanwhile, at $m_\phi \sim 10\,$MeV, the canonical freeze-out cross section into neutrinos-only can be larger than for annihilation into electrons-only, as DM annihilation affects the neutrino temperature stronger. At last, in each figure the curves with different branching ratios should gradually converge at $m_{\rm} \ge 30\,$MeV, when the freeze-out happens well before neutrino decoupling.

The results to other cases carry over in the following sense:  the DM  degrees of freedom, $g_{\rm DM}$, only enter logarithmically in the calculation of the freeze-out temperature, so the canonical thermal cross section is essentially the same for complex scalar and Dirac fermion DM.  In contrast, the self-conjugate cases, i.e., real scalar and Majorana fermion DM approximately require half the annihilation cross section compared to the non-self-conjugate cases to obtain the same final relic abundance; see App.~\ref{app:analyFO}.  Therefore, our Fig.~\ref{fig:thermal_xsec} is generalized to these other options of DM candidates with only mild losses in accuracy.

\begin{figure*}[t]
\begin{center}
\includegraphics[width=0.50\textwidth]{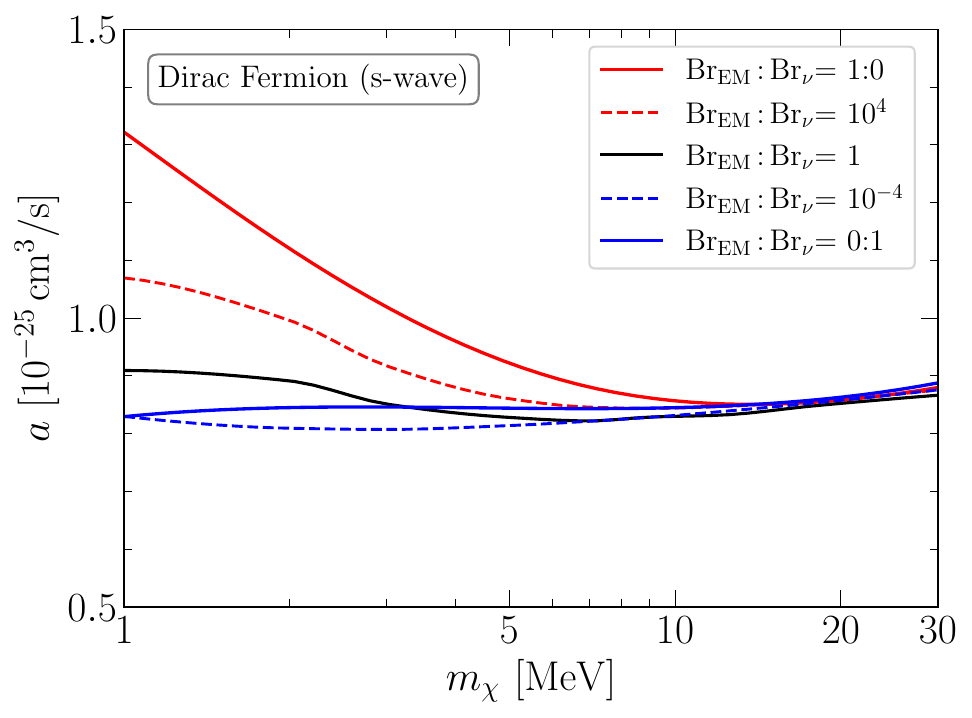}
\includegraphics[width=0.485\textwidth]{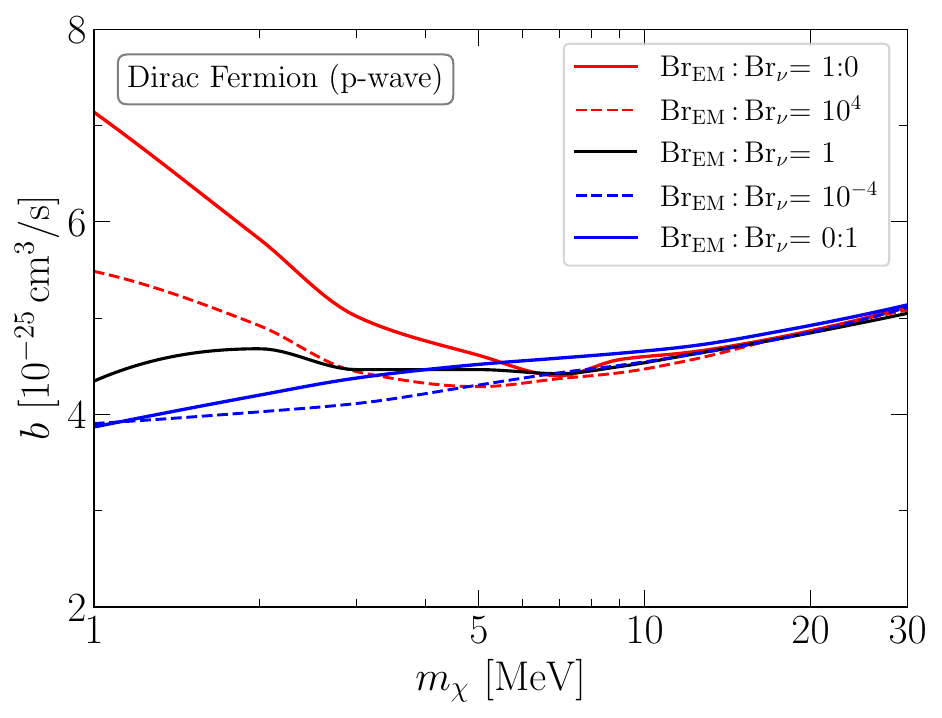}
\\
\includegraphics[width=0.49\textwidth]{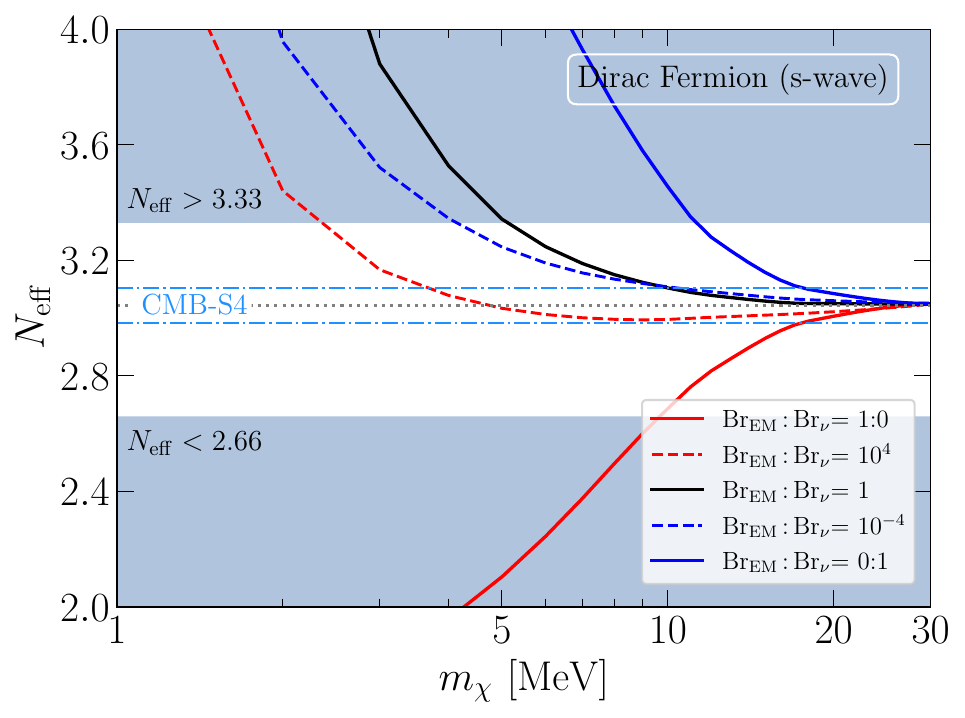}
\includegraphics[width=0.49\textwidth]{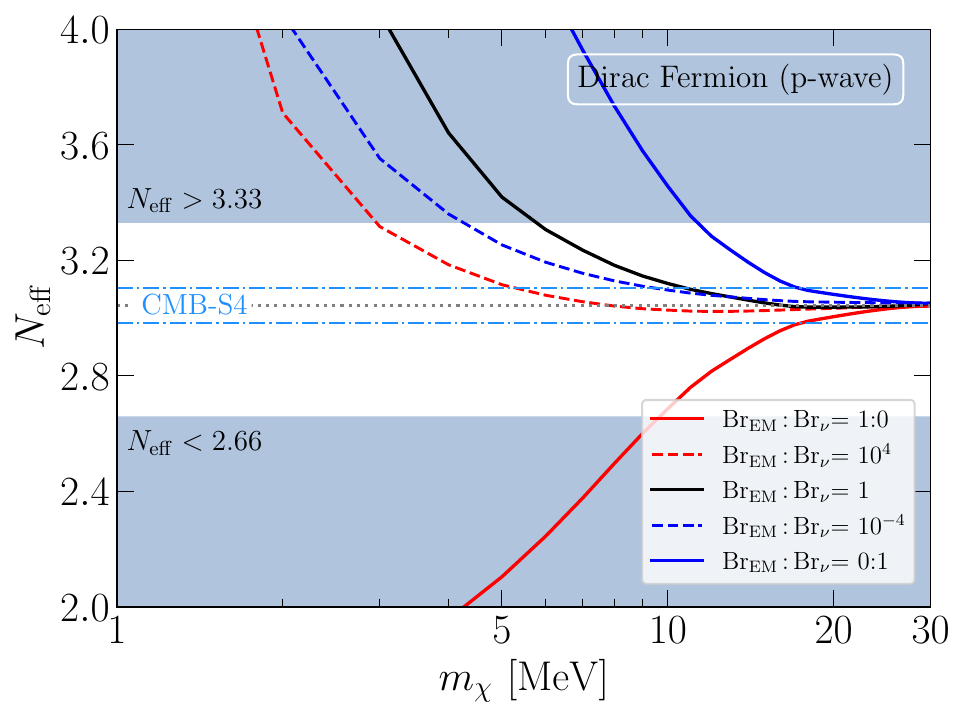}
\end{center}
\caption{Same as Figs.~\ref{fig:thermal_xsec} and \ref{fig:thermal_Neff} above, but for Dirac fermion DM. The left (right) panel gives and $s$-wave ($p$-wave) results.   
}
\label{fig:fermionicNX}
\end{figure*}

\subsection{\boldmath$N_{\rm eff}$ and minimal DM mass}

We now study the predictions for $N_{\rm eff}$ at the CMB epoch by adopting the established canonical values for the DM annihilation cross section.
To this end, we take  $N_{\rm eff}^{\rm SM}= 3.044$ as the standard cosmological history value (\textit{e.g.}, with conventional TeV-scale DM), consistent with our SM-only calculations. At 95\% C.L.~the
combination of Planck and baryonic acoustic oscillation (BAO) measurements yields $ 2.66 \leq N_{\rm eff} \leq 3.33$~\cite{Planck:2018vyg}.
In terms of the deviation from the standard value, 
\begin{align}
  \Delta N_{\rm eff} \equiv N_{\rm eff} - N_{\rm eff}^{\rm SM}\,,
\end{align}
the error-bar is expected to improve such that 
the expected sensitivity of the Simons Observatory reads $|\Delta N_{\rm eff}|\lesssim 0.1$~\cite{SimonsObservatory:2018koc}, and that of CMB-S4  $|\Delta N_{\rm eff}| \lesssim 0.06$~\cite{Abazajian:2019eic} when one assumes a standard cosmological history as the benchmark point.

The  results on $N_{\rm eff}$ are shown in Fig.~\ref{fig:thermal_Neff}. We find that for most of the branching ratios, the dark sector is able to transfer energy from the EM sector to the neutrino sector, increasing the value of $N_{\rm eff}$. Moreover, as we mentioned above, the effect is mostly sensitive to $\text{Br}_{\rm EM}\,\text{Br}_\nu$, so the two cases of $\text{Br}_{\rm EM}:\text{Br}_\nu = 10^{4}$ and $10^{-4}$ lead to similar results. 
Obviously, for $\text{Br}_{\rm EM} : \text{Br}_\nu= 1$ the EM- and $\nu$-sector are  most strongly connected via the DM ``agent'' so that this  branching ratio results in pronounced $N_{\rm eff}$ values, which are comparable to the case that DM only annihilates into neutrinos.  These features are in broad agreement with previous works~\cite{Escudero:2018mvt, Sabti:2019mhn}, while we also take into account the exact canonical cross section for each DM mass and the  sub-leading contribution of DM-SM elastic scattering.  The exact canonical cross section is two-to-three times the often assumed  value of one pico-barn which modifies the final bounds on the DM mass for $\text{Br}_{\rm EM}\,\text{Br}_{\nu}\neq 0$, depending on the exact values of branching ratio. In the cases of ${\text{Br}_{\rm EM}\,\text{Br}_\nu} = 0$, the final value of $N_{\rm eff}$ is simply decided by  entropy conservation after neutrino decoupling, regardless of the exact canonical cross section or whether it is $s$- or $p$-wave dominated. This can be seen from the similarity of the respective solid red and blue curves of the two panels of  Fig.~\ref{fig:thermal_Neff}.

We also investigate the canonical annihilation cross sections, as well as the associated CMB $N_{\rm eff}$ values, for Dirac fermion DM $\chi$ with both $s$-wave and $p$-wave annihilation. The results are shown in Fig.~\ref{fig:fermionicNX}.  While there are great similarities with the complex scalar DM case,  some differences exist at DM mass below 10\,MeV  due to the fact that a Dirac fermion has~4 degrees of freedom~(or $7/8\times 4 =  3.5$ effective bosonic degrees of freedom), which contributes to the total energy density of the Universe at $T_\gamma \sim$\,MeV. As a result, a slightly larger canonical annihilation cross section is needed to compensate for a larger Hubble rate. Its energy density also has a stronger impact on~$N_{\rm eff}$ in cases of ${\text{Br}_{\rm EM}\,\text{Br}_\nu}=0$, leading to stronger bounds on the DM mass. For non-negligible values of $\text{Br}_{\rm EM}\,\text{Br}_\nu$ where the final value of $N_{\rm eff}$ is mainly determined by the prolonged EM-$\nu$ kinetic equilibrium, the d.o.f. of the DM particle only mildly affects the bounds on the DM mass.

With those results at hand, Tab.~\ref{tab:minmass} lists the minimal thermal DM masses that are compatible with an otherwise standard cosmological history in the absence of additional particles and/or other ``model-building tricks.''  The limits vary from $m_\chi = 2\ \MeV$ to $11.2\ \MeV$ depending on the model and branching ratio. Only the first column with {\it no} annihilation into neutrino leads to a decrease in $N_{\rm eff}$, whereas all other cases increase $N_{\rm eff}$ from the standard value. The weakest limit (in terms of the lowest allowed DM mass) is attained for $\text{Br}_{\rm EM}: \text{Br}_\nu = 10^{4} $. As can also be observed the differences between complex scalar and Dirac fermion are rather mild, with $p$-wave annihilation leading to slightly stronger limits.

We close this discussion by commenting on the case of self-conjugate DM candidates. A Majorana fermion has two spin degrees of freedom and effectively $7/8\times 2$ bosonic degrees of freedom so that the case likely closely resembles the complex scalar one. This is also suggested in~\cite{Sabti:2019mhn}. Finally, a real scalar only has one bosonic degree of freedom and consequently relaxed bounds.

\begin{table}[t]
    \centering
    \begin{tabular}{l|lllll}
    \toprule
    \qquad\qquad\qquad $ {\rm Br}_{\rm EM}:{\rm Br_\nu}$  & 1:0 & $10^{4}$ & \,\,\,1 &   $10^{-4}$ & 0:1
       \\
\midrule
Complex scalar ($s$-wave) & 6.9  & 2.0 & 4.8 &  3.8 & 8.7\\
Complex scalar ($p$-wave) & 6.9 &  2.9 & 5.2 & 3.9 & 8.7\\
Dirac fermion ($s$-wave) & 9.5 & 2.5 & 5.0 & 4.1 & 11.2\\
Dirac fermion ($p$-wave) &  9.5 & 3.0 & 5.7 & 4.3 & 11.2  \\
\bottomrule
    \end{tabular}
    \caption{Minimal DM mass ($m_{\phi,\, \chi}/\MeV$) for various DM candidates and branching ratios into the EM and neutrino sector compatible with the 95\% C.L.~limit $2.66 \leq N_{\rm eff} \leq 3.33$~\cite{Planck:2018vyg}.}
    \label{tab:minmass}
\end{table}

\subsection{{Exemplary particle physics realizations}}
\label{sec:models}

Benchmark cases for MeV complex scalar DM~$\phi$ and Dirac fermion DM~$\chi$ interacting with charged leptons and neutrinos via a heavy mediator for both, $s$- and $p$-wave annihilation cases, are readily constructed. For example, 
for a (pseudo-)scalar mediator, the DM annihilation of a complex scalar $\phi$ is  $s$-wave, while for a vector mediator it is $p$-wave. For a fermionic $\chi$, a vector mediator can induce both, $s$- and $p$-wave annihilation.
Below, we discuss the $s$-wave annihilation of a complex scalar DM {which was used in obtaining the corresponding $s$-wave results above; the description of} the $p$-wave counterpart is given in our companion paper~\cite{Chu:2022xuh}. 
{The models for} $s$- and $p$-wave annihilating fermionic DM particle $\chi$ are provided in App.~\ref{app:DiracF}.

A simple realization for $s$-wave annihilation of complex scalar DM is the exchange of an intermediate real heavy pseudoscalar $A$ via the renormalizable interactions terms%
\begin{align}
\label{eq:scalar}
\mathcal{L}^{(A)}_{\rm int} &= 
 - i \sum_l y_l A (\bar{l} \gamma_5 l) -\mu_A A (\phi^* \phi) \,, 
\end{align}
where the sum runs over leptons $l=e,\nu_e, \dots$.\footnote{See \cite{Berlin:2014tja, Berlin:2015wwa} for concrete examples of pseudoscalar portals to a dark sector. Note that throughout this work SM neutrinos are  assumed to be of Majorana nature for pseudoscalar interactions. Additional pseudoscalar interactions of neutrinos can be induced via mixing with sterile neutrinos, see e.g. ~\cite{Archidiacono:2014nda, Fiorillo:2020jvy}.}
This leads to $s$-wave DM annihilation to neutrinos and electrons, and a velocity-dependent elastic scattering between DM and SM fermions.
At low energies the interaction is  described by the effective mass dimension-5 operator $i y_l (\phi^* \phi)(\bar l \gamma^5 l)/\Lambda$  with $\Lambda \equiv \mu_{A}/m^2_{A}$; the Yukawa couplings $y_l$ and trilinear coupling $\mu_{A}$ are taken as real.

The interactions in~\eqref{eq:scalar} give rise to annihilation processes such as $\phi^*\phi\leftrightarrow e^+ e^-$, $\phi^*\phi\leftrightarrow\bar \nu \nu  $ and elastic scattering processes such as $\phi e\leftrightarrow \phi e$, $\phi\nu\leftrightarrow \phi  \nu $. 
Varying the ratio $y_\nu /y_{e}$ then amounts to entertaining different combinations of  branching ratios of annihilation into neutrinos, ${\rm BR}_\nu$, and into electrons and/or photons, ${\rm BR}_{\rm EM}$ (here electrons).
The tree-level $\phi\phi^*$ annihilation cross sections in the non-relativistic limit read 
\begin{align}
\label{eq:annswave}
   \sigma_{e} v_M  &\simeq \dfrac{y_e^2}{4\pi \Lambda^2}\,  \left( 1 - \dfrac{m_e^2}{m_\phi^2} \right)^{1 / 2} \!\!\!\!+ \dfrac{y_e^2 m_f^2  v_{\rm rel}^2 }{32 \pi \Lambda^2 m_\phi^2  } \, \left( {1 - \dfrac{m_e^2}{m_\phi^2}} \right)^{-{1/2}} \, ,  \notag \\
\quad
  \sigma_{\nu} v_M  & \simeq \dfrac{y_\nu^2 }{8\pi \Lambda^2}  + {\mathcal O}(v_{\rm rel}^4)\, ,\notag 
\end{align}
where the $p$-wave component is subleading when $m_\phi \gg m_e$.
The Boltzmann equations that describe the number and energy densities are given in App.~\ref{app:be}. The collision terms that enter the evaluation for $s$-wave annihilation are calculated in App.~\ref{app:int_canonical} and for $p$-wave annihilation are listed in the companion paper~\cite{Chu:2022xuh}.

The parameter regions concerned for DM thermal freeze-out are  easily allowed by other existing bounds, if one takes the mediator mass to be a few to tens of GeV. Taking the electron-mediator interaction described by Eq.~\eqref{eq:scalar} as an example, intensity-frontier and neutrino experiments provide the strongest constraints on such a GeV-scale mediator, limiting $y_e/m_A$ to be below $10^{-3}$--$10^{-4}$\,GeV$^{-1}$, e.g., by searching for di-lepton resonances and/or neutrino-electron scattering signatures~\cite{BaBar:2014zli, Bilmis:2015lja}. On the other hand, the DM-mediator interaction in \eqref{eq:scalar} is best bounded by DM self-scattering. Requiring $\sigma_{\phi\phi\leftrightarrow \phi\phi}\lesssim 0.2$\,barn/GeV~\cite{Wittman:2017gxn, Eckert:2022qia}  leads to $\mu_A / m_A \lesssim 0.05(m_\phi/\text{MeV})^{3/4}$. Therefore the choice, say,  $\mu_A/m_A \lesssim 0.03$---compatible with all the constraints discussed above---is always allowed for the  DM mass range considered in this work. The situation is similar for other interactions adopted below, as long as the mediator mass is taken to be several GeV.

At last, general neutrino-DM interactions are much less  constrained experimentally,  like the annihilation channels discussed above. 
In fact, due to the weak interaction, a neutrino-philic mediator with a mass  above the GeV-scale is hardly probed by current experiments. Note that further constraints from neutrino non-standard interactions induced by the mediator may enter too. A detailed discussion on neutrino-DM/mediator interactions will be deferred until Sec.~\ref{sec:neutrino}. 

\section{Avoiding the \boldmath$N_{\rm eff}$ constraint}

As mentioned, the dark sector modifies $N_{\rm eff}$ with two competing effects: first, EM-$\nu$ kinetic equilibrium mediated by DM always increases $N_{\rm eff}$ with respect to $N^{\rm SM}_{\rm eff}$. 
{This is because in standard cosmology $T_\nu < T_\gamma$ after $e^\pm $ annihilation; a prolonged DM-mediated kinetic coupling of the EM and $\nu$ sectors will therefore raise the neutrino temperature relative to photons. }
Second, DM annihilation {\it after} EM-$\nu$ kinetic decoupling either increases or decreases $N_{\rm eff}$. It depends on the relative branching ratios  $\text{Br}_{\rm EM} : \text{Br}_\nu $, where a larger ratio heats up the EM sector better and therefore decreases $N_{\rm eff}$. For $\text{Br}_{\rm EM}: \text{Br}_\nu \gg 1$, a  parameter region exists where both effects  cancel. {To arrange for this cancellation, the kinetic decoupling needs to happen well before the DM energy density gets strongly Boltzmann-suppressed, requiring $\text{Br}_{\rm EM}  \text{Br}_\nu \ll 1$.
}

\begin{figure}[t]
\begin{center}
\includegraphics[width=0.5\textwidth]{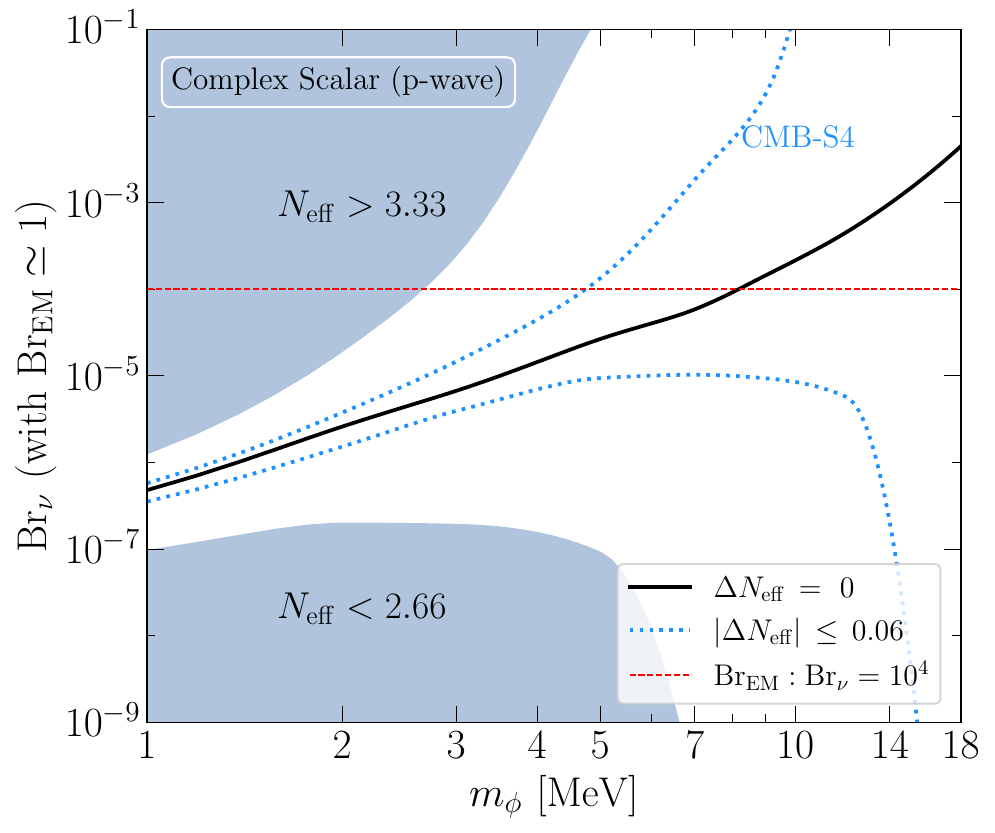}
\caption{Illustration of fine-tuned branching ratios that give small $N_{\rm eff}$  at recombination, as the increase of $N_{\rm eff}$ caused by delayed kinetic decoupling of  EM and neutrino sectors is canceled by the subsequent heating of the EM sector by DM annihilation after kinetic decoupling. The observed DM relic abundance fixes total cross sections. The increase on (D/H) remains below 2\% on the black line, saturating the value at $m_\chi = 1\ \MeV$. An elevated helium abundance excludes points along the black line for $m_\phi < 3\,\MeV$; see Tab.~\ref{tab:finetuned}.
}
\label{fig:SPoptimal}
\end{center}
\end{figure}

We illustrate this cancellation as the solid black line in Fig.~\ref{fig:SPoptimal} for complex scalar DM with $p$-wave annihilation.  The case of $s$-wave annihilation with similar parameters is already excluded by DM indirect searches.  Note that for even heavier DM,  with mass above $40\,$MeV, which freezes out well before $T_\gamma \sim m_e$ but after SM neutrino decoupling around $T_\gamma \sim 3$\,MeV, the optimal ratio of $\text{Br}_{\rm EM} : \text{Br}_\nu $ should eventually converge to $(g_\gamma + 7/8 g_e): 7/8 g_\nu = 21/22$.   The situation is very similar in the Dirac DM model, which would require even stronger fine-tuning to accommodate the bounds on $N_{\rm eff}$ as it has more effective d.o.f. than a complex scalar.  

\begin{figure}[t]
\begin{center}
\includegraphics[width=0.5\textwidth]{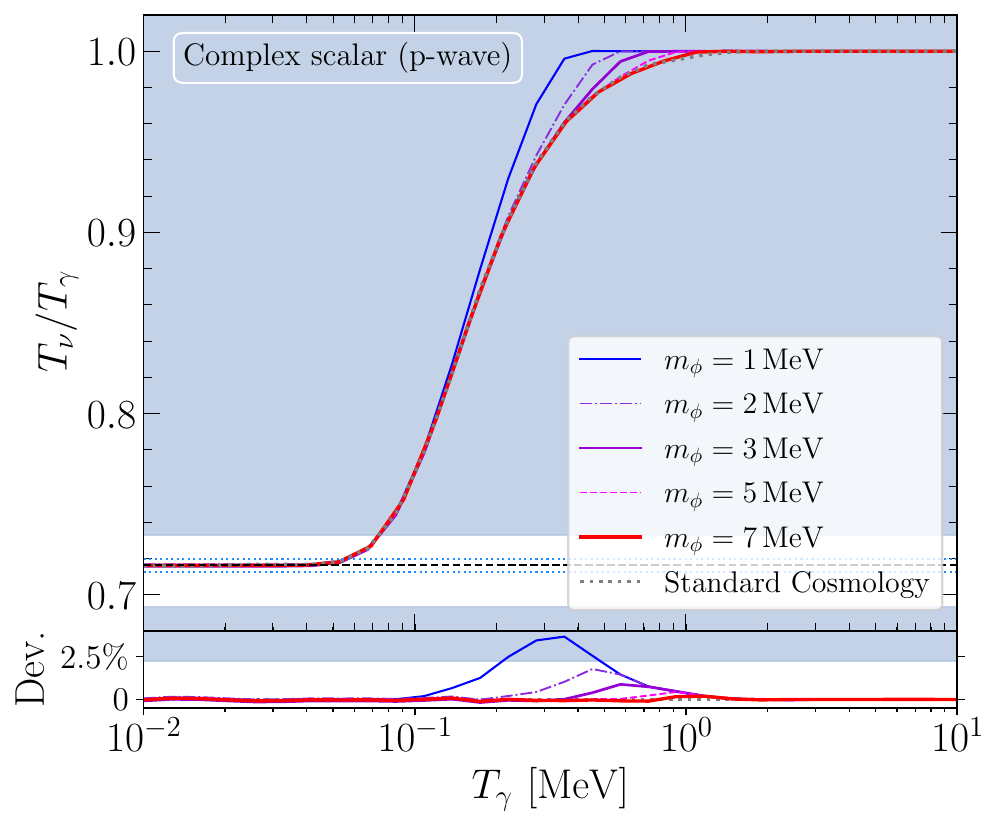}
\caption{Evolution of $T_\nu/T_\gamma$ for tuned parameters that result in negligible additional radiation at recombination with complex scalar DM and $p$-wave canonical freeze-out.  The black dashed horizontal lines correspond to its CMB value predicted by Standard cosmology, 0.7164, as well as the projected CMB-S4 sensitivity $|\Delta N_{\rm eff}| \le 0.06$. Blue-shaded regions indicate the current Planck$+$BAO bounds at recombination ($T_\gamma \sim 0.3\,$eV.) The lower panel shows the deviation of the $T_\nu/T_\gamma$ ratio of the respective cases in percent from the standard cosmological evolution. 
}
\label{fig:SPoptimalTEMP}
\end{center}
\end{figure}

For branching ratios that result in exact cancellation, we also provide the neutrino temperature evolution for scalar DM mass $m_{\rm \phi} = 1,\,3,\,5,\,7$\,MeV as a function the photon temperature in Fig.~\ref{fig:SPoptimalTEMP}. One observes from its lower panel that with the EM-neutrino kinetic coupling induced by DM ($T_\nu/T_\gamma = 1$), the neutrino sector is hotter w.r.t.~standard cosmology case. When the two sectors kinetically decouple ($T_\nu/T_\gamma \neq 1$), DM dominantly annihilates into the EM sector, reducing the ratio of $T_\nu/T_\gamma$.  %
Figures~\ref{fig:SPoptimal} and \ref{fig:SPoptimalTEMP}  show that one can always tune the branching ratios to satisfy the $N_{\rm eff}$ bounds from CMB.
However, MeV-scale thermal DM additionally contributes to the total energy-density of the Universe before its density gets Boltzmann suppressed. The latter modifies the predicted abundances of primordial light elements, and may thus receive constraints from BBN. To check this, we feed the neutrino- and photon-temperature as well as the DM energy density evolution into our BBN code {described in}~\cite{Pospelov:2010hj}, {including all necessary current updates of nuclear reaction rates following~\cite{Wiescher:2010eni}.}

Before including DM, we obtain a standard BBN (SBBN) deuterium abundance of  ${\rm D/H} = 2.49\times 10^{-5}$ and a helium mass fraction abundance of $Y_p = 0.2475$ in good agreement with literature~\cite{Pitrou:2018cgg}; 
a neutron lifetime of $879.5$~s~\cite{Serebrov:2017bzo} has been assumed. 
Recent measurements of the deuterium and helium abundances broadly agree with the SBBN predictions. Over the years, the helium values have ranged between $0.24\lesssim Y_p\lesssim 0.26$~ \cite{Izotov:2014fga,Cooke:2018qzw,Aver:2020fon,Kurichin:2021wlb}. The most aggressive 95\% C.L.~constraint $Y_p\leq 0.251$ results when employing recent observations with claimed small error bar, $Y_p = 0.247\pm 0.0020$~\cite{Kurichin:2021wlb}.
In turn, observations of deuterium abundances are now at the percent-level, ${\rm D/H} =(2.527 \pm 0.030)\times 10^{-5}$~\cite{Cooke:2017cwo,Cooke:2013cba}, allowing for a departure by  $\pm 2.4\%$ from its central value at 95\%~C.L.. Tab.~\ref{tab:finetuned} shows the results when DM is included with a branching ratio such that the $N_{\rm eff}$ constraint is evaded. We observe  that while the increase in (D/H) remains below or at $2\%$, the helium abundance increases with decreasing DM mass. This is the effect from the DM energy density itself and it excludes finely-tuned masses~$m_\chi\leq 2\ \MeV$.

\begin{table}[t]
    \centering
    \begin{tabular}{l|rrrrr}
    \toprule
    \qquad  (${\rm Br}_{\rm EM} \simeq 1$)   & \ $  {\rm Br}_{\nu} $ & $\Delta  (D/H)$ & $10 \times Y_p  $ & $\Delta Y_p  $ & viable?  \\
\midrule
SBBN &  -- & --        &  2.478 & -- & \cmark \\ 
$m_\chi=7\ \MeV$    & $5.9\times 10^{-5}$  & $-0.1\%$  & 2.479  & +0.1\%   & \cmark \\ 
$m_\chi=5\ \MeV$    & $2.7\times 10^{-5}$&  +0.1\% &  2.485  & +0.3\%   & \cmark \\ 
$m_\chi=3\ \MeV$    & $6.7\times 10^{-6}$ &  +0.5\% & 2.502 &  +1.0\% &  \cmark \\
$m_\chi=2\ \MeV$    & $2.6\times 10^{-6}$  &  +1.1\% & 2.525 &  +1.9\% &  \xmark  \\
$m_\chi=1\ \MeV$    & $4.8\times 10^{-7}$ &  +2.0\% & 2.568 &   +3.7\% & \xmark \\
         \bottomrule
    \end{tabular}
    \caption{Cosmological observables for the fine-tuned branching ratios that evade the $N_{\rm eff}$ constraint. The last column indicates the cosmological compatibility and it is driven by the helium abundance.}
    \label{tab:finetuned}
\end{table}

\section{Millicharged DM and its connection to 21~cm cosmology}
\label{sec:millicharged}

A well suited application of our three-sector treatment are MeV-scale millicharged particles, being entertained in a variety of contexts. An exciting prospect that has emerged in recent years is the potential of 21~cm cosmology as a probe of the Universe at high redshift $30 \lesssim z \lesssim 6$~\cite{Pritchard:2011xb}.
The emission of 21~cm radiation from  neutral hydrogen during that epoch is sensitive to the baryon temperature, which in turn may be altered by dark matter-baryon or dark matter-electron interactions~\cite{Tashiro:2014tsa,Munoz:2018pzp,Barkana:2018qrx}.
The current state of observations is controversial. Whereas the EDGES collaboration has claimed the observation of an absorption feature at redshifted 21~cm wavelength that is stronger than expected from standard cosmology~\cite{Bowman:2018yin}, the signal is contested by more recent observation by the SARAS3 instrument~\cite{Singh:2021mxo}. Nevertheless, 21~cm cosmology is a new window probing light dark sector physics. 

The required strong effective DM-baryon interaction cross section can be mediated by the Coulomb-like $v^{-4}$ velocity enhancement for which millicharged DM is a prime candidate.
Currently, light dark states with a millicharge between $5 \times 10^{-6} \lesssim \epsilon  \lesssim   10^{-4}(m_\chi/5\,{\rm MeV})^{0.6}$  are allowed by both the SLAC mQ experiment and SN cooling constraints~\cite{Kovetz:2018zan}.
Moreover, MeV-scale DM particles with $\epsilon \lesssim 10^{-5}$ become able to reach underground XENON detectors to trigger recoil signals, and thus be excluded~\cite{Liu:2019knx}. Similarly, to avoid the stringent constraint from the surface run of the SENSEI experiment, one needs even larger millicharges,  $\epsilon \gtrsim 8\times 10^{-5}$\cite{Crisler:2018gci, Wadekar:2019mpc}.\footnote{Note that the result of \cite{Wadekar:2019mpc} cannot be trivially re-scaled for a sub-leading MeV DM component, as mentioned in their paper. Besides, the upscattering of the incident DM flux from the solar corona still offers an avenue to probe this parameter space~\cite{An:2021qdl}.}   
The combination of mQ and direct detection experiments thus requires $m_\chi \gtrsim 3$\,MeV. 
On the flip side, to explain  EDGES (or more generally, to have an influence in 21~cm cosmology), heavier $\chi$~particles either need to make a larger portion of the total DM abundance, or one needs to introduce larger values of~$\epsilon$. Since  detailed investigations of CMB spectra demand a DM mass fraction below $0.4\%$ for the $\epsilon$ values discussed above~\cite{Kovetz:2018zan,Boddy:2018wzy, Buen-Abad:2021mvc,Nguyen:2021cnb}, there exist also upper bounds on $m_\chi$ values of interest. 
So, taken together, the parameter range of interest becomes $m_\chi$ from $3-30$\,MeV and $\epsilon$ of $(8-20)\times 10^{-5}$,  making up 0.1\%-0.4\% of the total DM abundance.

Nevertheless, the scenario remains challenged by early Universe cosmology: the sizable value of $\epsilon$ suggests that such MeV dark state, $\chi$, comes into thermal equilibrium with radiation during/before BBN. The annihilation into electron-positron pairs%
\footnote{Annihilation into a photon-pair is higher order in $Q$ and hence subleading for $m_\chi > m_e$.}
heats the EM sector relative to the $\nu$-sector. 
The parameter region of interest is then shown to be largely excluded because of a too low $N_{\rm eff}$ value~\cite{Creque-Sarbinowski:2019mcm}. 
With our developed method of treating three sectors during freeze-out, we may now check to which degree the above conclusion holds when the millicharged DM candidate $\chi$ instead dominantly annihilates into neutrinos via additional interactions. 

\subsection{Model and cross sections}

In order to make contact with preceding literature~\cite{Kovetz:2018zan}, we also take $\chi$ to be a Dirac fermion.
We may then consider the millicharge interaction being supplemented by a pseudoscalar mediator, $A$, similar to what was done in the previous section,%
\begin{equation}
\label{eq:millichargedL}
{\mathcal L}_{\rm int} =
-i\epsilon e A^\mu (\bar \chi \gamma_\mu \chi) - i  y_\nu  A  \sum_l (\bar \nu_l  \gamma_5 \nu_l) - i  y_A A  (\bar \chi \gamma_5 \chi) \,,
\end{equation}
where $A^\mu $ is the SM photon,  $e$ is the electric charge with $Q=\epsilon e$, and $l= e,\mu,\tau$. {As $\chi$ only makes up a sub-percent fraction of the DM abundance, the pseudoscalar coupling is not constrained by DM self-interaction bounds.}

The annihilation cross section of $\bar \chi \chi$ with squared center-of-mass energy $s$ into a pair of electrons, mediated by the millicharge, is given by 
\begin{align}
    \sigma_{e} v_M & = \dfrac{8\pi \epsilon^2 \alpha^2}{3s}  \left(1+{2m_e^2 \over s} \right) \left(1+{2m_\chi^2 \over s}\right) \sqrt{1 - \dfrac{4m_e^2}{s}} \notag \\
   & \simeq \dfrac{\pi \epsilon^2 \alpha^2}{m_\chi^2} \left(1 + \frac{m_e^2}{2m_\chi^2} \right)\sqrt{1-\frac{m_e^2}{m_\chi^2}}  . 
\end{align}
As a benchmark value we take $\epsilon =10^{-4}$ and $m_\chi =6$--$10$\,MeV,  which yields $ \sigma_e v_M  \simeq  (2\text{--}6) \times 10^{-25}\ \cm^3/{\rm s}$. This satisfies the CMB and indirect search bounds as $\chi$ only makes up sub-percent fraction of the total DM abundance.

For the dominant annihilation channel, into neutrinos, we fix the corresponding coupling so that $\chi$ and its antiparticles  amount for $0.2\%$ of the observed DM abundance. 
And as shown above,  the canonical freeze-out value for $m_\chi =6$--$10\,$MeV DM annihilation cross section into neutrinos is about $10^{-25}$\,cm$^3$/s. Therefore, we have the total non-relativistic annihilation cross section into neutrinos around $5\times 10^{-23}$\,cm$^3$/s,  translating into relative branching ratios of $\text{Br}_{\rm EM}:\text{Br}_\nu \sim 10^{-2}$. As the $s$-wave cross section for the new annihilation channel into a single neutrino flavor mediated by~$A$ is
\begin{equation}
    \sigma_{\nu} v_{M} 
     =
     \frac{y_A^2 y_\nu^2}{8\pi}\frac{s}{(m_A^2-s)^2}
     \simeq    \frac{y_A^2 y_\nu^2}{2\pi}\frac{m_\chi^2}{m_A^4}\,,
\end{equation}
thus the corresponding parameter set is given by
\begin{equation}\label{eq:MQnuXsec}
   3 \sigma_{\nu} v_{M} 
      \simeq 5\times 10^{-23}\,\text{cm}^3\ \text{s}^{-1} \,\left({y_A y_\nu   m_\chi\over 3\,{\rm MeV}}\right)^2   \left({ \text{GeV} \over m_A}\right)^4, 
\end{equation}
where the pre-factor 3 counts three neutrino flavors as final states. Applicable neutrino bounds for this parameter ballpark range are provided below in Sec.~\ref{sec:neutrino}.

Finally, we point out that the elastic scattering cross section between $\chi$ and neutrinos via each mediator reads
\begin{align}
    \sigma_{\chi \nu} =  & y^2_A y^2_\nu \dfrac{s}{48 \pi m_A^4}   \left(1-\dfrac{m_\chi^2}{s} \right)^2   \,, 
\end{align}
where we have assumed that $s \ll m^2_{A}$. 
Given that the typical squared center of mass energy scales as $(s-m_\chi^2) \sim m_\chi T_\nu$, this cross section has a temperature dependence such that the elastic scattering rate per $\chi$-particle steeply decreases with $T_\nu^5$, in analogy with neutrino-electron interactions.

\begin{figure}[t]
\begin{center}
\includegraphics[width=0.50\textwidth]{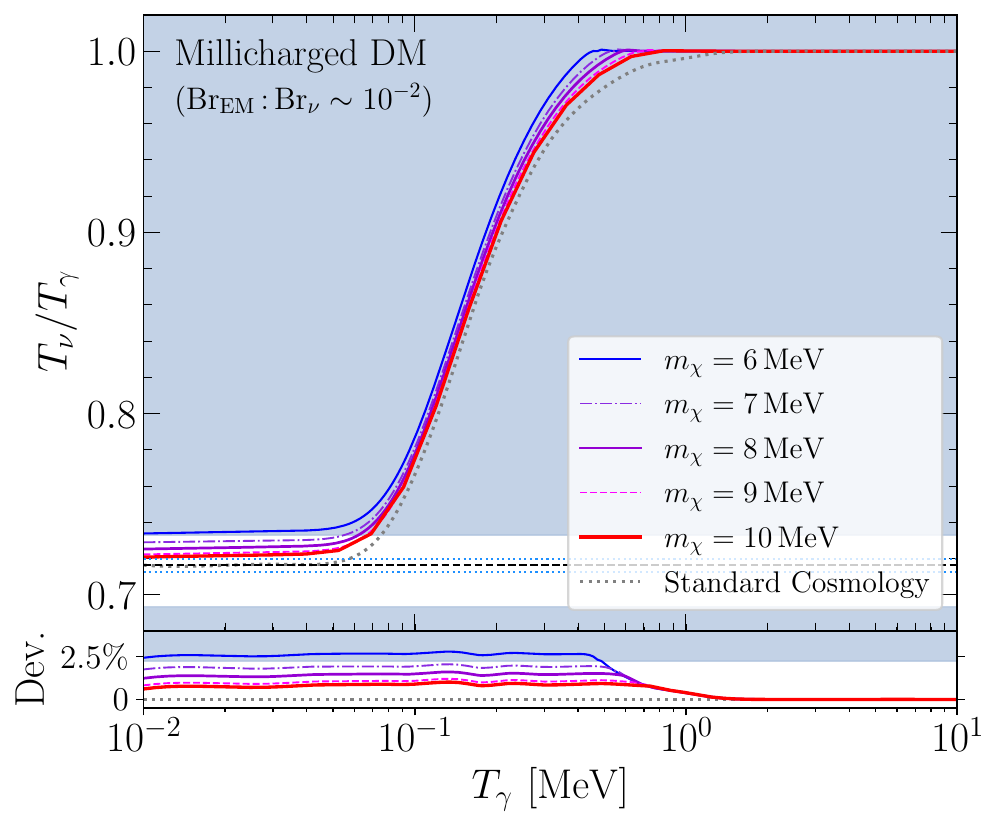}
\caption{
The evolution of $T_\nu/T_\gamma$ for millicharged DM as a function of photon temperature, with the same parameters as in Tab.~\ref{tab:bbn}. The black dashed horizontal lines correspond to its CMB value predicted by Standard cosmology and the projected CMB-S4 sensitivity, while blue-shaded regions indicate the current Planck$+$BAO bounds, same as Fig.~\ref{fig:SPoptimalTEMP}.
}
\label{fig:BBNTratio}
\end{center}
\end{figure}

\begin{table}[t]
    \centering
    \begin{tabular}{l|lrrrrr}
    \toprule
        & $N_{\rm eff}$ & $\Delta N_{\rm eff}$ & $\Delta  (D/H)$ & $Y_p \times 10 $ & $\Delta Y_p  $ & viable?  \\
\midrule
SBBN & 3.044        & --& --         &  2.478 & -- & \cmark \\ 
$m_\chi= 10\ \MeV $ & 3.119  & 0.075  & +1.0\% &  2.488  &  +0.4\%  & \cmark\\ 
$m_\chi=9\ \MeV$    &  3.171 & 0.127  & +1.5\% &  2.493  &  +0.6\%  & \cmark\\ 
$m_\chi=8\ \MeV $   & 3.193  & 0.149  & +1.8\% &  2.496  & +0.7\%   &  {\cmark}\\ 
$m_\chi=7\ \MeV$    & 3.268  & 0.224  & +2.8\% &  2.503  & +1.0\%   & {\bf ?} \\ 
$m_\chi=6\ \MeV$    & 3.352  & 0.308  & +3.8\% &  2.512  & +1.4\%   & \xmark \\ 
         \bottomrule
    \end{tabular}
    \caption{Cosmological observables for a millicharged DM particle with fractional abundance of $0.2\%$, with $\epsilon = 10^{-4}$ and Eq.~\eqref{eq:MQnuXsec}. The last column indicates the cosmological compatibility.}
    \label{tab:bbn}
\end{table}

\subsection{Cosmological constraints}

In order to test for the cosmological compatibility of the millicharged fractional DM scenarios annihilating into neutrinos, we now investigate the value of $N_{\rm eff}$ at CMB and compute the light element abundances from BBN.%
\footnote{Other works that consider the modifications of light element abundances from MeV-scale dark sectors include~\cite{Serpico:2004nm,Nollett:2013pwa,Nollett:2014lwa,Kawasaki:2015yya,Wilkinson:2016gsy,Depta:2019lbe,Berlin:2019pbq,Sabti:2019mhn,Berlin:2019pbq,Sabti:2021reh,Giovanetti:2021izc}.}
For the latter we directly feed the non-trivial evolution of the electromagnetic, neutrino, and dark matter densities $\rho_{\rm EM}$, $\rho_\nu$ and $\rho_\chi$ into a modified version of a nucleosynthesis code~\cite{Pospelov:2010hj}. We note in passing, that the neutrino annihilation products  that are being injected after neutrino-decoupling with $m_\chi$ initial energies are too few in number to induce non-thermal reactions such as $\bar \nu_e p \to e^+ n$ at a relevant level; see~\cite{Pospelov:2010cw} for a detailed discussion. 

Tab.~\ref{tab:bbn} summarizes the results of this analysis for $\epsilon=10^{-4}$ with cross section of Eq.~\eqref{eq:MQnuXsec}, and various DM masses at and below 10~MeV. For better comparison, the first line shows the results for a standard cosmological history. Our obtained value $N_{\rm eff}^{\rm SM} = 3.044$ is in agreement  with other recent literature results, $3.043-3.046$~\cite{deSalas:2016ztq, Bennett:2019ewm, EscuderoAbenza:2020cmq, Froustey:2020mcq, Akita:2020szl,  Bennett:2020zkv, Cielo:2023bqp}. 
The generally observed trend when {\it neutrino-annihilating} millicharged states are included is that for decreasing $m_\chi$, neutrino heating becomes pronounced, leading to elevated levels of $N_{\rm eff}$. In terms of $ \Delta N_{\rm eff}$ 
we observe shifts from $0.075$ to $0.308$, mostly compatible with the current 95\% C.L.~range inferred from Planck (see above).
With similar branching ratios, $\text{Br}_{\rm EM}:\text{Br}_\nu \sim 10^{2}$--$10^3$, and a total annihilation cross section at $3\times 10^{-26}$\,cm$^3$/s, Tab.~8 of~\cite{Escudero:2018mvt} obtains a lower bound  $m_\chi \ge  4.3\,$MeV 
from $ N_{\rm eff} \lesssim 3.33$. As shown in Fig.~\ref{fig:BBNTratio}, a much larger cross section of $5\times 10^{-23}$\,cm$^3$/s adopted here further delays the DM-induced decoupling between neutrino and EM sectors, increasing the bound to $m_\chi \gtrsim  6.1\,$MeV.

We now turn to the BBN results. 
Including $\chi$ into our calculation, we find an increasing trend with decreasing DM mass for both ${\rm D/H}$ and~$Y_p$. This effect is well known in the helium mass fraction and, within the considered $m_\chi$-range, the helium abundance is barely changed beyond the one per-cent level. In turn, for the lightest mass considered, $m_\chi \ = 6\ \MeV$, the deuteron abundance changes by 3.8\%, in tension with observations.
Given that the SBBN D/H prediction has a one percent uncertainty stemming from nuclear-rate uncertainties, we are not able to make a definite statement of the viability for $m_\chi = 7\ \MeV$. Still, the relative changes inform us that the scenario is on the verge of being best probed by~D/H, and upcoming improvements in the observations of $N_{\rm eff}$ will provide the definite test. Moreover, the 95\% C.L.~upper limit on $Y_p$ is only touched for the lightest mass $m_\chi = 6\,\MeV$. 

To sum up, for this scenario considered here to explain EDGES,  early Universe cosmology suggests a lower bound at $m_\chi \gtrsim 7\,\MeV$ at this moment. Furthermore, based on our discussion above, given the sizeable value of $\text{Br}_{\rm EM}\text{Br}_\nu $ adopted here, this bound is not expected to change much for a millicharged scalar case.

\subsection{Experimental bounds on neutrino interactions}
\label{sec:neutrino}

In the set-up given above, the presence of a neutrino-philic pseudo-scalar mediator induces neutrino self-interactions and neutrino-DM scattering, and thus can be constrained by experimental observations. In contrast, a sub-leading DM candidate as above is only poorly constrained by conventional DM searches, and a MeV dark particle with a charge $\epsilon = 10^{-4}$ is below the current sensitivity of intensity-frontier experiments; see e.g. \cite{Kovetz:2018zan, Wadekar:2019mpc, Foroughi-Abari:2020qar}. In this subsection, we consider relevant observations, with a benchmark mass of intermediate (pseudo-)scalar, $m_A \sim $\,GeV  and  $y_A y_\nu \simeq 0.3$.

Regarding such neutrino-DM interaction induced by $A$, the mean-free-path of a high-energy neutrino passing through the fractional DM medium can be estimated via 
\begin{equation}
    l_{\rm mfp} \sim 10^3\,\text{Gpc}\, \left({ 10^{-3}\over f_{\rm DM}}\right) \left({m_\chi \over {\rm MeV}}\right) \left({10^3\,  \text{GeV}^{-2} \over \sigma_{\chi \nu}} \right)\,,
\end{equation}
which means that a PeV neutrino freely travels in the Universe in our set-up~\cite{Choi:2019ixb}.
Moreover, recent $\mathcal O(100)$\,TeV neutrino observations in IceCube from Blazar TXS 0506+056 and active galaxy NGC 1068 could lead to 5-10 orders of magnitude stronger bounds, if such neutrinos were generated around supermassive black holes within dense DM spikes~\cite{Cline:2022qld, Ferrer:2022kei, Cline:2023tkp}. The conclusion of Ref.~\cite{Cline:2023tkp} can be re-scaled to give in our model
\begin{equation}
   y_A y_\nu \lesssim 10^2 \, \left({ 10^{-3}\over f_{\rm DM}}\right)^{1/2} \left({ m_A \over \text{GeV}} \right) \left({\,\text{MeV}\over m_\chi } \right)^{1/4} \,, 
\end{equation}
which applies to $m_\chi \gtrsim 1\,$MeV. This bound is also very weak since the spike should be truncated by efficient $\chi$-pair annihilation. 
Besides, neutrino-$\chi$ interaction inside a proto-neutron star may enhance dark particle capture, greatly  alleviating the SN cooling bounds on $\epsilon$, which is similar to self-trapping induced by $\chi$ self-interaction~\cite{Chang:2018rso, Sung:2021swd}. %

For a GeV mediator,  stronger bounds may come from neutrino self-interaction. There exists  $y_\nu \lesssim 0.3$ from observations of leptonic decay of mesons and heavy leptons, such as $K/D \to l\nu $ and $\tau \to l\nu \nu$~\cite{Pasquini:2015fjv, Berryman:2018ogk, Blinov:2019gcj}, which is stronger than the potential bounds from double beta decay~\cite{Deppisch:2020sqh}.  Strong neutrino self-interactions may also affect the neutrino-driven mechanism of SN explosions in a non-trivial way~\cite{Shalgar:2019rqe, Chang:2022aas,Fiorillo:2023ytr}. Therefore, future observations of SN neutrinos could further clarify the exact SN evolution, and thus probe this parameter region; see also e.g. \cite{Ko:2019ssx, Cerdeno:2023kqo}.
On the other hand, cosmological bounds on neutrino self-interaction are much weaker. For instance, CMB observations require neutrinos to free-stream at $z\le 10^4$ (assuming no recoupling), leading to $y^2_\nu/m_A^2 \le 0.1$\,MeV$^{-2}$~\cite{Brinckmann:2020bcn}. For a summary of neutrino self-interaction bounds, see a recent review \cite{Berryman:2022hds}.  As a result, the current limit on $y_\nu$ in turn requires $y_A$ values to be around, or slighter larger than, unity in our model.

At last we briefly comment on the contribution of new particles to the electromagnetic properties of neutrinos. In our set-up, neutrinos can  couple to photon via loops of intermediate pseudoscalar and millicharged particles. As the photon does not mix with a pseudoscalar, dimensional analysis suggests that in heavy scalar limit this can at most happen via dim-6 operators, such as anapole moment,  with coefficients of the order $(\epsilon e) m_\chi^2/m_A^4 \lesssim  10^{-8}$\,GeV$^{-2}(m_\chi/\text{10\,MeV})^2 (\text{GeV}/m_A)^4$, or even smaller as $(\epsilon e)   m_\nu^2/m_A^4$. Such benchmark values of this model are well below the current bounds.  For instance, the upper bound on  neutrino charge radius is, $\langle r^2 \rangle \lesssim 10^{-6}$\,GeV$^{-2}$ (about $10^{-33}$\,cm$^2$)~\cite{ParticleDataGroup:2020ssz, MammenAbraham:2023psg}, for the parameter region of our concern.   

{In summary, one may amend the millicharged DM model by neutrino interactions with a GeV-mass mediator and induce the necessary branching ratio into neutrinos without facing new constraints that could not be evaded on the account of taking $y_A\sim O(1)$.}

\section{Conclusions}
\label{sec:conclusions}

In this work, we explored the prediction of DM candidates with a mass below 30~MeV and which come into thermal equilibrium with the SM. Such states undergo thermal freeze-out close to or during the epoch of neutrino-decoupling, affecting the SM predictions of the ratio of neutrino to photon temperatures, or, equivalently, of~$N_{\rm eff}$. When DM annihilates into both neutrinos and electrons/photons, an accurate prediction of $N_{\rm eff}$ as well as of the relic density, necessitate the simultaneous solution of the coupled three-sector system: DM, neutrinos, and the EM sector. The methodology for achieving this in a fully self-consistent way that takes into account energy transfer into the various sectors from both annihilation and elastic scattering  was developed in our preceding work~\cite{Chu:2022xuh}. 

Here we utilize this formulation to derive the thermal values of the annihilation cross section that yield the correct relic abundance for an exemplary set of branching ratios into neutrinos and electrons for $s$- and $p$-wave annihilation. For example, an $s$-wave annihilating complex scalar $\phi$ with $m_\phi =1~\MeV$ requires a total annihilation cross section of $a\simeq 7.5\times 10^{-26}\ \cm^3/{\rm s}$ for dominant annihilation into neutrinos, ${\rm Br_{EM}}/{\rm Br}_\nu < 10^{-4} $, and an annihilation cross section equal or larger than $10^{-25}\ \cm^3/{\rm s}$ for dominant annihilation into electrons, ${\rm Br_{EM}}/{\rm Br}_\nu > 10^{4} $. For $m_\phi = 15\ \MeV$ all cases converge to approximately   $a\simeq8\times 10^{-26}\ \cm^3/{\rm s}$, in agreement with previous investigations. For the $p$-wave annihilating case and $m_\phi =1~\MeV$ we obtain $b\simeq 3.4\times 10^{-26}\ \cm^3/{\rm s}$ for ${\rm Br_{EM}}/{\rm Br}_\nu < 10^{-4}$, and  $b \geq 5\times 10^{-26}\ \cm^3/{\rm s}$ for dominant annihilation into electrons, ${\rm Br_{EM}}/{\rm Br}_\nu > 10^{4}$. Small differences among the cases pertain to higher masses---an effect that traces back to the non-trivial temperature evolution in the dark sector; at   $m_\phi = 20\ \MeV$ all cases require  $b\simeq 4.5\times 10^{-26}\ \cm^3/{\rm s}$. The equipartitioned cases with $ {\rm Br}_{\rm EM}/{\rm Br_\nu} =1 $ lie in between the above values. For a Dirac fermion the observed trends as a function of mass are similar, but thermal cross sections mildly differ from the complex scalar case.

Using the thermal cross section values, we are then in the unique position to obtain a precision determination of $N_{\rm eff}$ as a function of $m_\phi$ and contrast it with current and future CMB-inferred bounds and projections. Exclusive annihilation into neutrinos (electrons) raises (lowers) $N_{\rm eff}$, excluding $m_{\phi} < 8 (6)\ \MeV$ for complex scalars and $m_{\chi} < 11 (9)\ \MeV$ for Dirac fermions
at 95\%~C.L.~ for both $s$- and $p$-wave annihilation from Planck measurements. Those limits are lowered when annihilation proceeds simultaneously into neutrinos and the EM sector. For ${\rm Br_{EM}}/{\rm Br}_\nu = 10^4 $ {\it and}  $10^{-4}$ an elevated $N_{\rm eff}$ excludes DM mass below $2\ \MeV$ {\it and}  below $ 4\ \MeV$, respectively, for both $s$-wave and $p$-wave annihilation and both model cases. Precise values for all cases are shown in Figs.~\ref{fig:thermal_Neff} and~\ref{fig:fermionicNX} and Tab.~\ref{tab:minmass}.

We also explore the possibility of a fine-tuned parameter region in ${\rm Br_{EM}}/{\rm Br}_\nu \gg 1$  with dominant annihilation into electrons where the minimum DM mass value can be lowered further as the heating of the photon and neutrino baths proceeds such that $N_{\rm eff}$ remains almost unchanged. For $p$-wave annihilation this presents a loophole to entertain lower-mass thermal DM; for $s$-wave DM any significant annihilation into electrons is already excluded from energy injection during the CMB epoch. For complex scalar DM and $p$-wave annihilation, we establish a required branching ratio into neutrinos of 
${\rm Br}_\nu \simeq 5\times 10^{-7}$ for $m_{\phi} = 1\ \MeV$ to $10^{-4}$ for $m_{\phi} = 10\ \MeV$ that remain unchallenged by current and future CMB $N_{\rm eff}$ measurements; see Fig.~\ref{fig:SPoptimal}. To complete the cosmic viability test, we also feed the non-standard evolution of photon and neutrino temperatures as well as the DM energy density into a BBN code and calculate the light element abundance yields. We find that $m_{\rm DM} < 3\, \MeV$ is excluded by an elevated helium abundance, see Tab.~\ref{tab:finetuned}. Finally, we complement those studies by presenting  particle physics realizations where we summarize other relevant observational and experimental constraints. 

In a second part, as a further case-study, we focus on the predictions of MeV-scale millicharged Dirac fermions $\chi$ in the mass-coupling regime where it affects the predictions of the global cosmological 21~cm signal through its thermal coupling to baryons at the cosmic dawn era. Because of other cosmological constraints, such states must have a sub-percent level fractional abundance. We show that it is possible to have a consistent thermal history of such particles when $\chi$ is being supplied with additional neutrino interactions that dominate the DM freeze-out. This  allows $\chi$ itself to deplete in number sufficiently. We compute $N_{\rm eff}$ and light element yields in the modified thermal history and find that a narrow mass-window situated around $10\,\MeV$ survives all current observational and experimental tests. Nevertheless, this window may become firmly closed by direct detection experiments soon.

Calculations of the thermal DM relic density are standard repertoire when entertaining dark sector models. Yet, the lowest mass range for simple thermal relics, $1-30$~MeV, considered on the brink of being compatible with cosmology, has not yet been studied at the appropriate level of rigor. In this work we have closed this gap. A further application of the three-sector approach may be considered the annihilation of $d$-wave annihilating DM or the decay of MeV-scale particles during the non-trivial epoch of neutrino decoupling and electron annihilation.

\begin{figure*}[t]
\begin{center}
\includegraphics[width=0.49\textwidth]{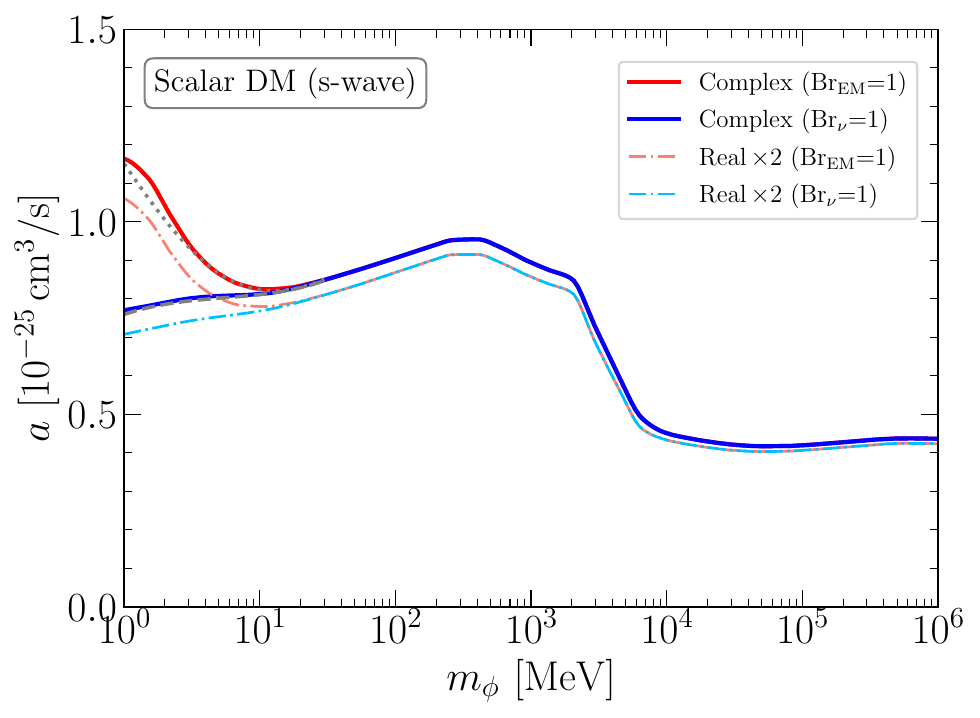}~
\includegraphics[width=0.49\textwidth]{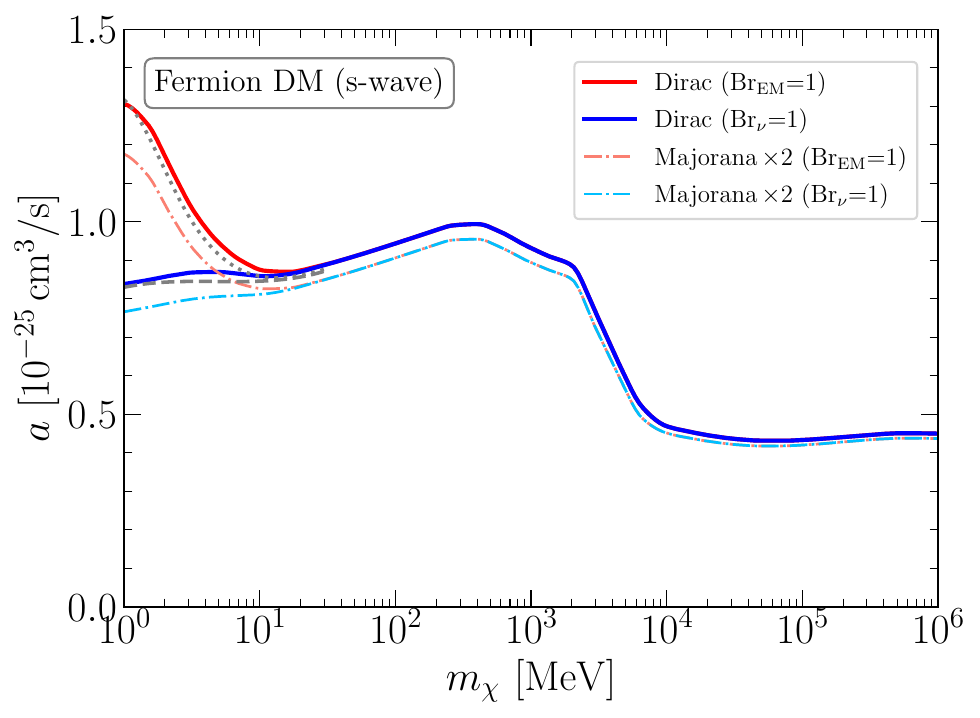}\\
\includegraphics[width=0.49\textwidth]{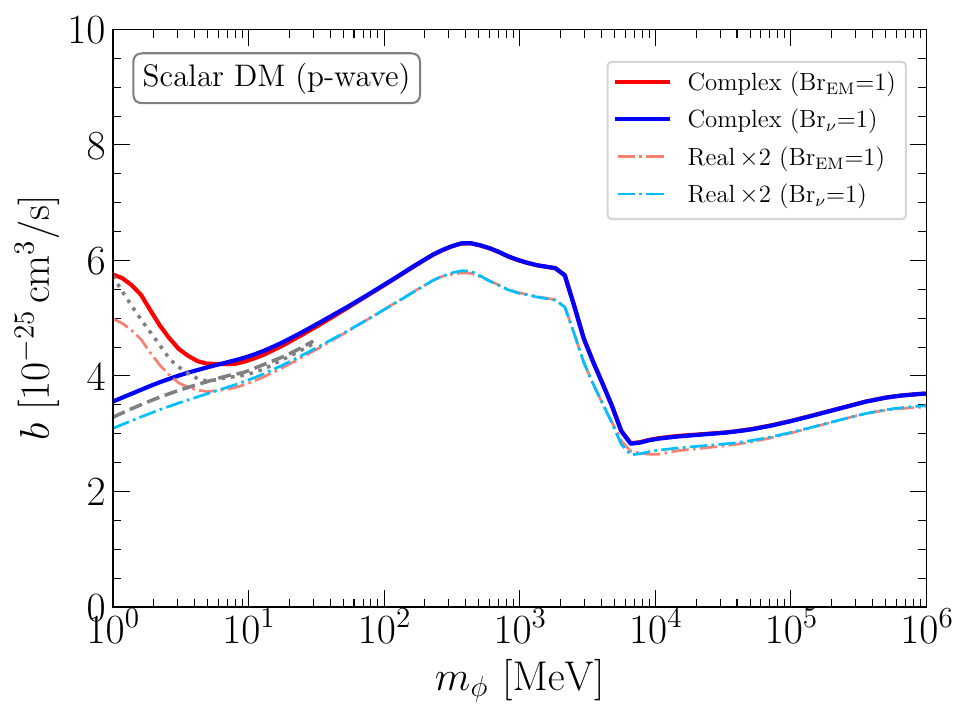}~
\includegraphics[width=0.49\textwidth]{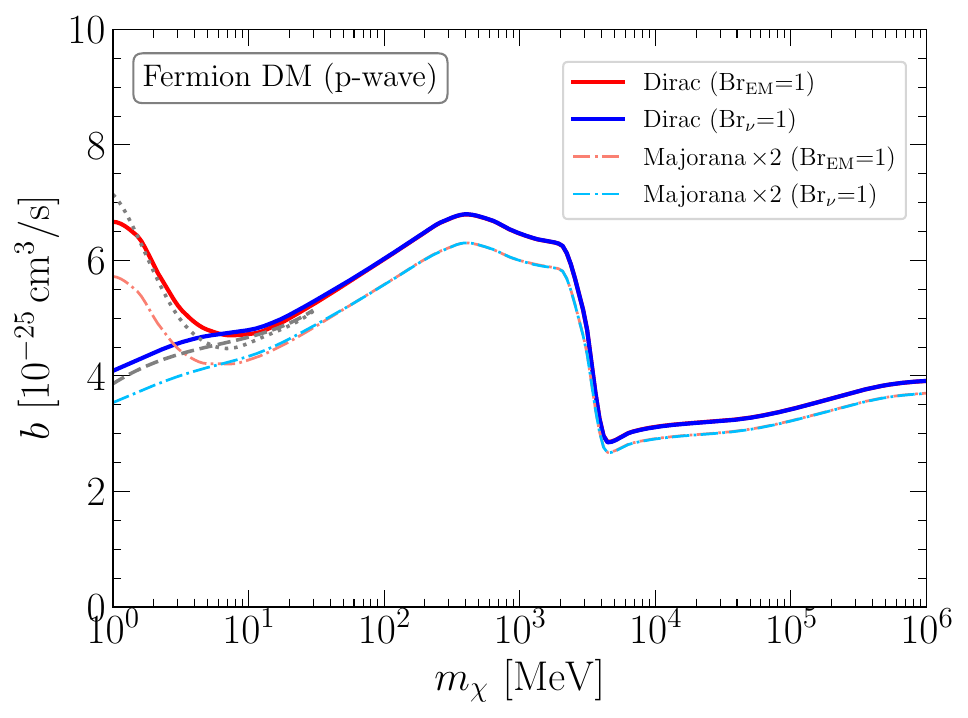}
\end{center}
\caption{Canonical cross section needed to obtain the observed abundance for scalar DM $\phi$ (left  panels) and fermion DM $ \chi$ (right panels) in the case of $s$-wave annihilation (top) and  $p$-wave annihilation (bottom) using the semi-analytic approach detailed in App.~\ref{app:analyFO}. Red lines are for exclusive pair annihilation into the EM sector, while blue lines are for exclusive annihilation into neutrinos. The y-axes give the parameter values of $\langle \sigma_{\rm ann} v\rangle  =  a + b(6T/m_{\phi,\chi}) $. Dot-dashed lines are  canonical values for self-conjugate DM, re-scaled by a factor of~2, as  self-conjugate cases need weaker annihilation. The gray dotted (dashed) lines give the exact numerical result for electron-only ($\nu$-only) annihilation obtained in the main text. }
\label{fig:FOcanonical}
\end{figure*}

\vspace{.3cm}

\paragraph*{Acknowledgments.}
We thank Jui-Lin Kuo for collaboration in the early phases of this work and for providing an initial version of the numerical code.
This work was supported by the FWF Austrian Science Fund research teams grant STRONG-DM (FG1) and by the U.S.~National Science Foundation (NSF) Theoretical Physics Program, Grant PHY-1915005.
Funded/Co-funded by the European Union (ERC, NLO-DM, 101044443). Views and opinions expressed are however those of the author(s) only and do not necessarily reflect those of the European Union or the European Research Council. Neither the European Union nor the granting authority can be held responsible for them. This work was also supported by the Research Network Quantum Aspects of Spacetime (TURIS). The computational results presented were in part obtained using the Vienna ``CLIP cluster''.

\appendix

\section{Semi-analytical freeze-out solutions}
\label{app:analyFO}

In this appendix, we also provide a semi-analytical solution to freeze-out in the case that DM with number density $n$ couples exclusively to the EM {\it or} the $\nu$ sector, i.e.~${\rm Br}_\nu {\rm Br}_{\rm EM} =0$. The derivation largely follows  the literature, such as Ref.~\cite{Gondolo:1990dk, Steigman:2012nb}, and the results are summarized in Fig.~\ref{fig:FOcanonical}. The aim is to illustrate the solution works for both $T  = T_\gamma$ and  $T   = T_\nu$,  by simply re-defining the entropy/energy degrees of freedom.

As we are only concerned with non-relativistic freeze-out, we assume DM follows a thermal evolution governed by the temperature of the sector it couples  to, $T_\gamma$ or $T_\nu$.  For this (semi-)analytical solutions in Fig.~\ref{fig:FOcanonical}, we shall assume a sudden neutrino decoupling at $T_\gamma = T_\nu = 2\,$MeV. 
One can start by defining an ``effective'' total entropy, $\hS= a^3 (2\pi^2 g_{\mathcal S} T^3 /45) \equiv a^3 \hs$,  which is conserved during the whole epoch through the appropriate choice of $g_\hs(T)$. We emphasize again that  $T$ can be the temperature shared by DM and its coupled sector, either $T_\gamma$ or $T_\nu$. In each case, one  solves the Boltzmann equation  
\begin{eqnarray}
    {d Y \over d t}   =  - \hs \langle \sigma_{\rm ann} v\rangle (Y^2 - Y^2_{\rm eq})\,. 
\end{eqnarray}

At the initial stage of the freeze-out evolution,  the DM abundance follows its thermal value  $Y_{\rm eq}$, up to a small correction, with 
\begin{equation}
    Y_{\rm eq} = {n\over \hs} \simeq 0.1447\,\left( {g_{\rm DM} \over g_{\hs} }\right) x^{1.5} e^{-x}\,. 
\end{equation}

We then follow a common convention and define the freeze-out point through $Y(x_{\rm fo}) = (1+c) Y_{\rm eq}(x_{\rm fo})$ with a constant $c$. It implies that at $x= x_{\rm fo}$  
\begin{equation}
        \left. {d \ln Y_{\rm eq} \over d \ln  x}\right|_{x_{\rm fo}}  =  - Y_{eq}  {\hs  \langle \sigma_{\rm ann} v\rangle  \over H} \left(1-  {1\over 3}{d \ln g_\hs \over d \ln x} \right)  (2+c)c \, \notag
\end{equation}
holds. By re-writing the equation as
\begin{equation}
      e^x =  - 0.1447\,\left( {g_{\rm DM} \over g_{\hs} }\right) x^{1.5} { {\hs  \langle \sigma_{\rm ann} v\rangle }    \over  H } { \left(1-  {1\over 3}{d \ln g_\hs \over d \ln x} \right)   (2+c)c  \over   {d \ln Y_{\rm eq} \over d \ln  x} } \,\, \notag
\end{equation}
together with ${d \ln Y_{\rm eq} / d \ln  x} = - (x-1.5 + d \ln g_\hs / d \ln x )$, we refer to the R.H.S. of the equation above as ${\mathcal F}(x)$, and re-write the whole equation as $ e^x = {\mathcal F}(x) $. 
A numerical solution to this equation can be obtained iteratively, %
\begin{equation}
    x_{\rm fo} = \ln {\mathcal F}|_{\{x \to \ln {\mathcal F}|_{[x \to ... ]} \}}\,,
\end{equation}
where as initial value of $x$ for MeV DM freeze-out one may choose 10-20.

Once the freeze-out point is reached, $Y$  becomes increasingly  larger than $ Y_{\rm eq}(x \gg x_{\rm fo})$, but  smaller than $ Y_{\rm eq}(x=x_{\rm fo})$, allowing for another approximation of the Boltzmann equation which may be cast in the form
\begin{equation}\label{eq:Boltzmann}
         {d Y \over Y^2}  =  - {\hs  \langle \sigma_{\rm ann} v\rangle  \over H x} \left(1-  {1\over 3}{d \ln g_\hs \over d \ln x} \right) dx \,.
\end{equation}
Integrating both sides from $x= x_{\rm fo}$ to  $x= \infty$ gives
\begin{equation}
      Y_{\infty} \simeq  \left[  \int^{\infty}_{x_{\rm fo} } {\hs  \langle \sigma_{\rm ann} v\rangle  \over H x} \left(1-  {1\over 3}{d \ln g_\hs \over d \ln x} \right) dx  \right]^{-1} \,.
\end{equation}

For the analytical solution of DM annihilating into  neutrinos only, one simply replaces $x \equiv  m_{\rm }/T_\gamma$  by $ x_\nu \equiv m_{\rm }/T_\nu$, and re-writes the functions of $g_{*}$ and $g_{\hs}$ with respect to $T_{\nu}$.%
\footnote{As usual, $g_*$ and $g_{\hs}$ are  defined as $g_*= \sum_b g_{b}(T_b/T)^4 + 7/8 \sum_f g_{f}(T_f/T)^4$ and $g_{\hs}= \sum_b  g_{b}(T_b/T)^3 + 7/8 \sum_f g_{f}(T_f/T)^3$ with $g_{b}$ and $g_{f}$ being the active bosonic (fermionic) relativistic degrees of freedom.} Since the neutrino sector is not heated up in electron annihilation, it generally makes the DM particles cooler than $T_{\gamma}$, and thus requires less DM annihilation to yield the observed DM abundance, as illustrated in the figure. Besides, for DM masses well above MeV,  DM particles do not contribute much to the total energy density of the Universe, thus the canonical cross sections become almost independent of the spin of DM particles.

The solutions of this semi-analytical method are shown as colored lines in Fig.~\ref{fig:FOcanonical}, which agrees with our full numerical results (gray lines) studied in the main text. Note that for the figure we have taken $c=0.4$ for both $s$-wave and $p$-wave annihilation, as it fits the exact numerical results of MeV-scale DM better.  If following \cite{Kolb:1988aj}, adopting $c=0.75$ for $p$-wave cases would increase the canonical cross sections by $5\%$--$10\%$.  For fermionic DM above $20$\,MeV, it is also in good agreement with previous literature values, see  e.g.~\cite{Steigman:2012nb, Bringmann:2020mgx}.  
Similar to~\cite{Bringmann:2020mgx}, we observe that the ratio in $p$-wave canonical cross sections between non-self-conjugate and self-conjugate cases visibly deviates from $2$ (lower panel of Fig.~\ref{fig:FOcanonical}) even for TeV DM. 
At last, we emphasize that this semi-analytical solution always assumes the kinetic equilibrium between DM and the SM sector it couples to. In contrast, the full numerical results adopt concrete particle models, and typically have DM kinetically decoupled from all radiation at $x\sim 100$. The latter thus requires a slightly larger canonical annihilation cross sections in cases of velocity-suppressed annihilation.

\section{Boltzmann equations}
\label{app:be}

Here we summarize the definitions and notation explained in further detail in the preceding methodology paper~\cite{Chu:2022xuh}.
The  evolution of number and energy densities, $n_i$ and $\rho_i$ of species $i$ is given by the Boltzmann equations,%
\begin{align}
       &  \dfrac{ \partial n_i}{\partial t} + 3Hn_i \equiv \dfrac{\delta n_i}{\delta t}\,, \label{eq:BZMn} \\
   & \dfrac{\partial  \rho_i}{\partial t} + 3H(\rho_i + P_i)\equiv \dfrac{\delta \rho_i}{\delta t}   \,,\label{eq:BZMrho}
\end{align}
with $P_i$ being the pressure density.
The right-hand-side terms $\delta n_i/\delta t $  and $ \delta \rho_i/\delta t $ are, respectively, given in terms of  sums over all contributing annihilation channels and all two-body processes (annihilation and scattering), 
\begin{align}
   & \dfrac{\delta n_i}{\delta t} =   g_i \int \dfrac{d^3 p_i}{(2\pi)^3 E_i}\,\sum_{\rm ann.} C[f_i] \,\Delta n  \,, \\
   &  \dfrac{\delta \rho_i}{\delta t}  =   g_i \int \dfrac{d^3 p_i}{(2\pi)^3 E_i}\,  \, \sum_{\rm all} C[f_i] \,\delta E \,, 
\end{align}
 where $C[f_i]$ is the collision term as it appears in the unintegrated Boltzmann equation for the phase space distribution function~$f_i$ of species~$i$; $\Delta n$ and $\delta E$ are the number and  energy exchanged in the process under question. In practice, we take $\Delta n =\pm 2$ for pair creation/annihilation. %
 For dark matter, which is assumed to develop a non-vanishing chemical potential after becoming non-relativistic,\footnote{This assumption may not be proper if the DM sector only couples to neutrinos. While this relies on exact parameters of the model, such as the mediator mass, we further assume that when ${\rm DM DM} \leftrightarrow \nu \nu$ is sufficient, $4\nu  \leftrightarrow  2\nu$ via dark sector mediators is sufficient too, resulting in $\mu_{\rm DM} = \mu_\nu =0$ during this period.} its momentum distribution function is approximated as 
 \begin{equation}
     f_{\rm DM} (E_{\rm DM}) \simeq  {e^{\mu_{\rm DM} /T_{\rm DM} }\over e^{E_{\rm DM} /T_{\rm DM} } \pm 1}\,, 
 \end{equation}
where ``DM'' is set as $\phi$  ($\chi$) for scalar (fermionic) dark matter candidate. For neutrinos, which are always relativistic and only obtain  tiny chemical potentials, there exists,  to the first order
  \begin{equation}
     f_{\nu } (E_\nu) \simeq {1 \over e^{E_\nu  /T_\nu  } + 1} + {\mu_\nu/T_\nu  \over e^{E_\nu  /T_\nu  } +e^{ - E_\nu  /T_\nu } + 2} \,. 
 \end{equation}
At last, the electromagnetic sector has no chemical potential above keV. 
For more details and uncertainty discussions, see~\cite{Chu:2022xuh}.
 
Our methodology of solving the three-sector system then involves factorizing the collision integrals in products of functions of two (normalized) chemical potentials $\tilde \mu_i = \mu_i/T_i$ and three temperatures~$T_i$, 
\begin{align}
\label{dn}
    \dfrac{\delta n_i}{\delta t} &= \sum\limits_{i\neq j} a_{ij} \,\beta_{ij}(\tmu_i, \tmu_j) \,\gamma_{ij} (T_i, T_j) \,, \\
    \label{drho}
     \dfrac{\delta \rho_i}{\delta t} &= \sum\limits_{i\neq j} b_{ij} \,\beta_{ij}(\tmu_i, \tmu_j) \,\zeta_{ij} (T_i, T_j) \,.
\end{align}
Here,  $a_{ij}=\pm 1 $ and $b_{ij}=\pm 1$, depending on the process under question, 
and $\beta_{ij}$ are functions of initial state chemical potentials such as $e^{\tmu_i+\tmu_j}$ or  $\tmu_i e^{\tmu_j}$.
Elastic scattering processes only enter in $\zeta_{ij}$ as particle number is conserved.

Analytical approximations for $\gamma_{ij} $ and $\zeta_{ij} $ for the charged scalar $s$-wave case are given in App.~\ref{app:int_canonical}, for the $p$-wave annihilation case they are found in the preceding work~\cite{Chu:2022xuh} and for the millicharged scenario they are given in App.~\ref{app:EDGES}.

\section{collision terms for $s$-wave scalar DM}
\label{app:int_canonical}

Here we provide the interaction rates for the pseudo-scalar mediated complex scalar DM model given in~\eqref{eq:scalar}. 
In the concrete expressions below,  the d.o.f. of all initial states have been summarized to give the physical number/energy-exchange between two sectors with fixed temperatures.

For the annihilation process $ee \leftrightarrow \phi\phi$, we obtain
\begin{align}
  \left(\dfrac{\delta n}{\delta t} \right)_{ee \leftrightarrow \phi\phi} &= \beta_{ee \leftrightarrow \phi\phi} \gamma_{ee \leftrightarrow \phi\phi}\,, \\
    \left(\dfrac{\delta \rho}{\delta t} \right)_{ee \leftrightarrow \phi\phi} &= \beta_{ee \leftrightarrow \phi\phi} \zeta_{ee \leftrightarrow \phi\phi} \,,
\end{align}
with $\beta_{ee \leftrightarrow \phi\phi} = e^{2\tmu_e}$ and 
\begin{align*}
\gamma_{ee \leftrightarrow \phi\phi} &= \dfrac{g_e g_{\bar e}}{(2\pi)^4} \int \dfrac{ds dE_+ dE_-}{2} \, f_e^{\rm eq} f_e^{\rm eq} \sigma_{ee \rightarrow \phi\phi} \mathcal{F}_{12} \\
&\times \left[ (1-\Delta_{\rm ann}) + \Delta_{\rm ann} (1- \beta_{\rm ann})\right] \,,\\
\zeta_{ee \leftrightarrow \phi\phi} &= \dfrac{g_e g_{\bar e}}{(2\pi)^4} \int\dfrac{ds dE_+ dE_-}{2} \, f_e^{\rm eq} f_e^{\rm eq} \sigma_{ee \rightarrow \phi\phi} \mathcal{F}_{12} \\
&\times E_+ \left[ (1-\Delta_{\rm ann}) + \Delta_{\rm ann} (1- \beta_{\rm ann})\right] \,, 
\end{align*}
where $g_e\equiv  g_{\bar e} = 2$. Here, the electron equilibrium distribution functions are given at vanishing chemical potential,
\begin{align}
    f_e^{\rm eq} \simeq \dfrac{1}{e^{E_e/T_\gamma} + 1 }\quad (T_\gamma \gtrsim m_e/20)\,.
\end{align}
The flux factor for a reaction of the type $1+2 \leftrightarrow 3+4$ is given by, 
$ \mathcal{F}_{12} = {\sqrt{\lambda(s,m_1^2,m_2^2)}}/{2}\,,
$ 
with $\lambda(a,b,c) = a^2 +b^2 +c^2 -2 (ab+ac+bc)$ being the triangle function. As usual, $s= (p_1+p_2)^2$ is the squared center-of-mass (CM) energy and $E_\pm = E_1 \pm E_2$. 
Finally, the weights are  defined as  $\Delta_{\rm ann.} \equiv e^{ (T_1^{-1} - T_3^{-1}) E_+}$ and $\beta_{\rm ann.} \equiv e^{2(\tmu_3 - \tmu_1)} =  e^{2(\tmu_4- \tmu_2)}$; here we have for the subscripts $1\widehat{=}	 e$ and $3 \widehat{=}	\phi$.
For the pseudoscalar induced annihilation, we obtain  
\begin{align}
  \sigma_{ee \rightarrow \phi\phi} = \dfrac{y_e^2 }{32\pi  \Lambda^2}\,{(1-4m_\phi^2/s)^{1/2}\over (1-4m_e^2/s)^{1/2}}\,.
\end{align}
Note that we always set the heavier particles as final states in annihilation cross sections, as we often neglect quantum statistics of final states to simplify the numerical computation; see \cite{Chu:2022xuh} for quantitative discussions and how the full statistics can be further included.

For the elastic scattering $\phi e \leftrightarrow \phi e$ we obtain,
\begin{align}
    \left(\dfrac{\delta \rho}{\delta t} \right)_{\phi e \leftrightarrow \phi e} &= \beta_{\phi e \leftrightarrow \phi e}  \zeta_{\phi e \leftrightarrow \phi e} \,,
\end{align}
with $\beta_{\phi e \leftrightarrow \phi e}  = e^{\tmu_\phi + \tmu_e}$ and 
\begin{align*}
\zeta_{\phi e \leftrightarrow \phi e} &= \dfrac{(g_\phi +g_{\phi^*} )( g_e + g_{\bar e}) }{(2\pi)^4} \int dE_1  dE_2  ds  dt \, f_\phi^{\rm eq}  f_e^{\rm eq}  \nonumber \\&\times \dfrac{d\sigma_{\phi e \rightarrow \phi e}}{dt} \, \mathcal{F}_{12} \langle \Delta_{\rm sca}  \delta E \rangle\,,
\end{align*}
where $g_\phi \equiv  g_{\phi^*} =1$ and  $f_\phi^{\rm eq}$ is the $\phi$ equilibrium distribution function at vanishing chemical potential,
\begin{align}
    f_\phi^{\rm eq} = \frac{1}{e^{E_\phi/T_\phi} -1} .
\end{align}
In the collision integral, $\langle \Delta_{\rm scatt.} \delta E \rangle$ is the energy transfer per scattering, averaged over the azimuthal angle in the CM frame; the explicit expression is found in~\cite{Chu:2022xuh}. The integration region of $t$ is given by $\left[ -\lambda(s,m_1^2,m_2^2)/s, 0 \right]$.

For the  pseudoscalar mediator, the differential cross section of elastic scattering is 
\begin{align}
\dfrac{d\sigma_{\phi e \rightarrow \phi e}}{dt} = \dfrac{- y_e^2  t}{16\pi \Lambda^2[m_e^4 - 2m_e^2 (m_\phi^2 +s) + (m_\phi^2 -s)^2]}\,.
\end{align}
Throughout this work, we set $d\sigma_{\rm scatt.}/dt$ positive, with $t$ being negative as defined above.

Turning to neutrino interactions with $\phi$. To avoid confusions,  only interaction terms with left-handed neutrinos are included in this work, so we can safely negelct right-handed neutrinos. Consequently,  we take $g_\nu\equiv g_{\bar \nu}=1$ for each neutrino flavor, distinguishing a left-handed neutrino from a right-handed anti-neutrino; $N_g=3$ gives the number of neutrino families.

Now, the collision term for neutrino pair annihilation, $\nu \nu \leftrightarrow \phi\phi$, is given by 
\begin{align}
  \left(\dfrac{\delta n}{\delta t} \right)_{\nu\nu \leftrightarrow \phi\phi} &= \gamma^0_{\nu\nu \leftrightarrow \phi\phi} +\beta^1_{\nu\nu \leftrightarrow \phi\phi} \gamma^1_{\nu\nu \leftrightarrow \phi\phi} \,, \\
    \left(\dfrac{\delta \rho}{\delta t} \right)_{\nu\nu \leftrightarrow \phi\phi} &= \zeta^0_{\nu\nu \leftrightarrow \phi\phi}+\beta^1_{\nu\nu \leftrightarrow \phi\phi}  \zeta^1_{\nu\nu \leftrightarrow \phi\phi} \,.
\end{align}
Here, the superscripts $0$ and $1$ signify an expansion in the neutrino chemical potential.  The collision integrals are given by 
\begin{align*}
\gamma^{0(1)}_{\nu\nu \leftrightarrow \phi\phi} (T) &= \dfrac{g_\nu g_{\bar \nu} N_g}{(2\pi)^4} \int \dfrac{dsdE_+ dE_-}{2} \, f_\nu^{\rm eq} f_\nu^{\rm eq (, 1)} \sigma_{\nu\nu \rightarrow \phi\phi} \mathcal{F}_{12} \\&\times \left[ (1-\Delta_{\rm ann}) + \Delta_{\rm ann} (1- \beta_{\rm ann})\right]  \,,\\
\zeta^{0(1)}_{\nu\nu \leftrightarrow \phi\phi} (T) &= \dfrac{g_\nu g_{\bar \nu}  N_g}{(2\pi)^4} \int \dfrac{dsdE_+ dE_-}{2} \, f_\nu^{\rm eq} f_\nu^{\rm eq (, 1)} \sigma_{\nu\nu \rightarrow \phi\phi} \mathcal{F}_{12} \\&\times E_+ \left[ (1-\Delta_{\rm ann}) + \Delta_{\rm ann} (1- \beta_{\rm ann})\right] \,. 
\end{align*}
where we set ~\cite{Chu:2022xuh}
\begin{align}
f^{\rm eq}_\nu (E_\nu/T_\nu) &= \frac{1}{ e^{E_\nu/T_\nu }  + 1}\,, \\
f^{\rm eq,1}_\nu (E_\nu/T_\nu) &= \frac{1}{ e^{E_\nu/T_\nu } +e^{- E_\nu/T_\nu} + 2}\,,\label{eq:fnuone}
\end{align}
and $\beta^1_{\nu\nu \leftrightarrow \phi\phi}  = 2 \tmu_\nu$. We have neglected the second-order corrections that are proportional to $f^{\rm eq,(1)}f^{\rm eq,(1)}$. 
For the  pseudoscalar-mediated annihilation it reads,
\begin{align}
  \sigma_{\nu\nu  \rightarrow \phi\phi} = \dfrac{y_\nu^2 }{16\pi   \Lambda^2}\,{(1-4m_\phi^2/s)^{1/2}}\,,
\end{align}
This cross section is a factor of 2 larger than the case of electron annihilation (in the limit $m_e\to 0$), the reason being that here we take each chiral SM neutrino as one degree of freedom, as stated above.

For the elastic scattering $\phi \nu \leftrightarrow \phi \nu$, we obtain
\begin{align}
    \left(\dfrac{\delta \rho}{\delta t} \right)_{\phi \nu \leftrightarrow \phi \nu} &= \beta^0_{\phi \nu \leftrightarrow \phi \nu}  \zeta^0_{\phi \nu \leftrightarrow \phi \nu} + \beta^1_{\phi \nu \leftrightarrow \phi \nu}  \zeta^1_{\phi \nu \leftrightarrow \phi \nu}\,,
\end{align} 
with $\beta^0_{\phi \nu \leftrightarrow \phi \nu} = e^{\tmu_\phi}$, $\beta^1_{\phi \nu \leftrightarrow \phi \nu} = e^{\tmu_\phi} \tmu_\nu$ and 
\begin{align*}
    \zeta^{0(1)}_{\phi \nu \leftrightarrow \phi \nu} &= \dfrac{(g_\phi +g_{\phi^*} ) (g_\nu+g_{\bar \nu}) N_g }{(2\pi)^4} \int dE_1 dE_2 ds dt \, f_\phi^{\rm eq} f_\nu^{\rm eq, (1)}  \\&\times \dfrac{d\sigma_{\phi \nu \rightarrow \phi \nu }}{dt} \, \mathcal{F}_{12} \langle \Delta_{\rm sca} \delta E \rangle\, .
\end{align*}
The differential cross section for the pseudoscalar-mediated $\phi$ interaction  reads 
\begin{align}
   \dfrac{ d\sigma_{\phi \nu \rightarrow \phi \nu }}{dt} = \dfrac{- y_\nu^2 t}{16\pi\Lambda^2 (m_\phi^2 -s)^2}\,.
\end{align} 
 And the cross section is the same for anti-neutrinos (same below). 

\section{collision terms for $p$-wave scalar DM}

The $p$-wave case follows the dark gauge boson mediator model in our preceding methodology paper~\cite{Chu:2022xuh}, and thus we further assume that electrons and neutrinos may carry different dark charges, labeled as $y_e$ and $y_\nu$. The (differential) cross sections, averaging over initial states and summing up final states are given as follows. 

For DM-neutrino interactions, there are  pair  creation/annihilation 
\begin{align}
  \sigma_{\nu\nu \rightarrow \phi\phi} = y_\nu^2 \dfrac{(s-4m_\phi^2)   }{24\pi  \Lambda_{Z'}^4} \left( {1-{4m_\phi^2 \over s}}\right)^{1/2}  \,,
\end{align}
and elastic scattering
\begin{align}
   \dfrac{ d\sigma_{\phi \nu \rightarrow \phi \nu }}{dt} =  y_\nu^2 \dfrac{(m_\phi^2 -s)^2 +st }{4 \pi \Lambda_{Z'}^4 (m_\phi^2 -s)^2}\,,
\end{align} 
per neutrino species. For DM-electron interactions,  there are also creation/annihilation 
\begin{align}
  \sigma_{ee \rightarrow \phi\phi} = y_e^2 \dfrac{(s-4m_\phi^2)  (s+2m_e^2)}{48\pi \Lambda_{Z'}^4  \, s} \left( \dfrac{ {s-{4m_\phi^2}} }{  {s-{4m_e^2}} } \right)^{1/2} \,.
\end{align}
and elastic scattering
\begin{align}
\dfrac{d\sigma_{\phi e \rightarrow \phi e}}{dt} = y_e^2  \dfrac{(m_e^2 +m_\phi^2 -s)^2 + t (s-m_e^2)}{4\pi \Lambda_{Z'}^4 [ m_e^4 - 2 m_e^2 (m_\phi^2 +s) +(m_\phi^2 -s )^2 ]}\,.
\end{align}

\section{$s$-wave and $p$-wave Dirac DM}
\label{app:DiracF}

In the case of $s$-wave Dirac DM, we take the simplest case 
\begin{equation}
{\mathcal L}^{\rm S} =\sum_l {y_l  \over \Lambda_{Z'}^2  }\left(\bar \chi \gamma^\mu       \chi \right )   \left(\bar l     \gamma_\mu  l \right) \,,
\end{equation}
where we assume that charged leptons and neutrinos can have different dark charges, being similar to the $p$-wave scalar case above. Then the associated four (differential) cross sections are
\begin{align}
 \sigma_{e e \rightarrow \chi \chi} &=  y_e^2 \dfrac{(s+2m_e^2)(s+ 2m_\chi^2)}{12\pi \Lambda_{Z'}^4 \, s }  \left( \dfrac{ {s-{4m_\chi^2}} }{  {s-{4m_e^2}} } \right)^{1/2}\,,\\
  \dfrac{ d\sigma_{\chi e \rightarrow \chi e}}{dt} & = y_e^2  \dfrac{2(s-m_e^2 -m_\chi^2)^2 + 2 s t +t^2 }{8 \pi \Lambda_{Z'}^4 [ (s - m_\chi^2)^2 + m_e^4 - 2 m_e^2 (s+ m_\chi^2)]} \,,
\end{align} 
for electrons, and for each neutrino species 
\begin{align}
 \sigma_{\nu \nu \rightarrow \chi \chi } &=  y_\nu^2  \dfrac{(s+ 2m_\chi^2)}{6\pi \Lambda_Z^4} \left( {1-{4m_\chi^2 \over s}}\right)^{1/2}\,,\\
  \dfrac{ d\sigma_{\chi \nu \rightarrow \chi \nu }}{dt} &  = y_\nu^2  \dfrac{ 2 (s -   m_\chi^2 )^2 + 2 t(s -m_\chi^2) + t^2 }{8 \pi \Lambda_{Z'}^4   (s - m_\chi^2)^2 } \,.
\end{align}

In the case of  the $p$-wave annihilation,  for simplicity, we only adopt  the interaction term 
\begin{equation}
{\mathcal L}^{\rm P} = \sum_l{\tilde y_l    \over \Lambda_{Z'}^2  }\left(\bar \chi \gamma^\mu  \gamma^5    \chi \right )   \left(\bar l     \gamma_\mu  l \right) \,,
\end{equation}
for which the associated four (differential) cross sections read
\begin{align}
 \sigma_{e e \rightarrow \chi \chi} &= \tilde  y_e^2 \dfrac{(s+2m_e^2)(s-  4m_\chi^2)}{12\pi \Lambda_{Z'}^4 s } \left( \dfrac{ {s-{4m_\chi^2}} }{  {s-{4m_e^2}} } \right)^{1/2}   \,,\\
  \dfrac{ d\sigma_{\chi e \rightarrow \chi e}}{dt} & =\tilde  y_e^2  \dfrac{2(s-m_e^2 -m_\chi^2)^2 + (2 s    - 4m_e^2  + t)t- 8m_e^2 m_\chi^2  }{8 \pi \Lambda_{Z'}^4 [ (s - m_\chi^2)^2 + m_e^4 - 2 m_e^2 (s+ m_\chi^2)]} \,,
\end{align} 
for electrons, and for each neutrino species 
\begin{align}
 \sigma_{\nu \nu \rightarrow \chi \chi } &= \tilde  y_\nu^2  \dfrac{(s -  4 m_\chi^2)}{6\pi \Lambda_{Z'}^4   } \left( {1-{4m_\chi^2 \over s}}\right)^{1/2}  \,,\\
  \dfrac{ d\sigma_{\chi \nu \rightarrow \chi \nu }}{dt} &  = \tilde  y_\nu^2  \dfrac{ 2 (s -   m_\chi^2 )^2 + 2 t(s -m_\chi^2) + t^2 }{8 \pi \Lambda_{Z'}^4   (s - m_\chi^2)^2 } \,.
\end{align}

\section{Millicharged Dirac DM coupled to neutrino-philic pseudoscalar}
\label{app:EDGES}

We now turn to fermionic DM models, for instance,  various collision terms predicted from the millicharged dark states supplied with neutrino interactions~\eqref{eq:millichargedL}.

For the annihilation channel $ee \leftrightarrow \chi \chi$ we write,
\begin{align}
  \sigma_{ee \rightarrow  \chi\chi} = \dfrac{4\pi \alpha^2 \epsilon^2 (s+ 2m_e^2) (s+2m_\chi^2)  }{3s^3  } \left( \dfrac{ {s-{4m_\chi^2}} }{  {s-{4m_e^2}} } \right)^{1/2}\,.
\end{align}

For the elastic scattering $\chi e \leftrightarrow  \chi e$, the interaction rates can be expressed as 
\begin{align}
\dfrac{d\sigma_{\chi e \rightarrow \chi e}}{dt} =  \dfrac{2\pi\alpha^2 \epsilon^2\left[2(m_e^2 +m_\chi^2 -s)^2 +2st +t^2 \right]}{(t-m_{\gamma,{\rm eff}}^2)^2 \left[m_e^4 -2 m_e^2 (m_\chi^2 +s) + (m_\chi^2 -s)^2 \right]}\,.
\end{align}
The effective photon mass $m_{\gamma,{\rm eff}}$ induced by finite-temperature effects has been introduced to avoid potential collinear divergence,  given by~\cite{Mirizzi:2009iz}
\begin{align}
m_{\gamma,{\rm eff}}^2 (T_\gamma) = \left\lbrace 
\begin{matrix*}[l]
4\pi \alpha n_{e} /m_e\,, & \quad(T_\gamma \ll m_e)\\
2\alpha \pi T_\gamma^2/3\,, &  \quad(T_\gamma \gg m_e)
\end{matrix*} \right.
\end{align}
where $n_e$ is the number density of electrons plus positrons.

Turning to the neutrino interactions  mediated by a pseudoscalar, we get
the (differential) cross sections as 
\begin{align}
  \sigma_{\nu\nu \rightarrow \chi\chi} = y_A^2 y_\nu^2 \dfrac{s}{8\pi \Lambda^4} \left( {1-{4m_\chi^2 \over s}}\right)^{1/2} \,,
\end{align}
and 
\begin{align}
 { d \sigma_{\chi\nu \rightarrow \chi\nu} \over dt}  = y_A^2 y_\nu^2  \dfrac{t (t-2m_\chi^2 )}{16\pi \Lambda^4 (s-m_\chi^2)^2 }  \,,
\end{align}
which applies to each SM neutrino species.

\bibliography{refs}

\end{document}